\documentclass[12pt]{article}

\usepackage{hyperref}
\usepackage{graphicx}
\usepackage{amsmath}
\usepackage{amsfonts}
\usepackage{amssymb}
\usepackage{eufrak}
\usepackage{mathrsfs}

\def\arXiv#1{\href{http://xxx.lanl.gov/hep-th/abs/#1}{#1}}

\input epsf.sty

\def\be{\begin{equation}}
\def\ee{\end{equation}}
\def\ba{\begin{eqnarray}}
\def\ea{\end{eqnarray}}
\def\mc{\mathcal}

\def\appendix{{\section*{Appendix}}\let\appendix\section%
        {\setcounter{section}{0}
        \gdef\thesection{\Alph{section}}}\section}

\begin{document}

\thispagestyle{empty}
\def\thefootnote{\fnsymbol{footnote}}
\begin{flushright}
 RUNHETC-2003-30 \\
 hep-th/0311177 \\
 \end{flushright}
\vskip 0.5cm

\begin{center}\LARGE
{\bf Nonsinglet Sector of $c=1$ Matrix Model and $2D$ Black Hole}\\
\end{center}

\vskip 1.0cm

\begin{center}
{\centerline{Satabhisa Dasgupta\footnote{{\tt
satavisa@physics.rutgers.edu}}}}

\vskip 0.5cm

{\it Department of Physics and Astronomy, Rutgers University,
\\ Piscataway, NJ 08854, U.S.A.}
\end{center}

\vskip 0.5cm

\begin{center}
{\centerline{Tathagata Dasgupta\footnote{{\tt
dasgupta@physics.nyu.edu}}}}

\vskip 0.5cm

{\it New York University Physics Department,
\\ 4 washington Place, New York, NY 10003, U.S.A.}
\end{center}

\vskip 1.0cm

Extending our recent work (\arXiv{\tt hep-th/0310106})
we study the nonsinglet sector of $c=1$ matrix model
by renormalization group analysis for a gauged matrix
quantum mechanics on circle with an appropriate gauge
breaking term to incorporate the effect of world-sheet
vortices. The flow equations indicate BKT phase transition
around the self-dual radius and the nontrivial fixed points
of the flow  exhibit black hole like phases for a range of
temperatures beyond the
self-dual point. One class of fixed point interpolate  between
$c=1$ for $R > 1$ and $c=0$ as $R \to 0$ via black
hole phase that emerges after the phase transition.
The other two classes of nontrivial fixed points also
develop  black hole like behavior beyond $R=1$.
From a thermodynamic study of the free
energy obtained from the Callan-Symanzik equations we show that
all these unstable phases do have negative specific heat. The
thermodynamic quantities indicate that the system does undergo a
first order phase transition near the Hagedorn
temperature, around which the new phase is formed, and
exhibits one loop finite energy correction to the Hagedorn density
of states. The flow equations also suggest a deformation of the
target space geometry through a running of the compactification
radius where the scale is given by the dilaton. Remarkably there is 
a regime where cyclic flow is observed. 

\vfill
\setcounter{footnote}{0}
\def\thefootnote{\arabic{footnote}}
\newpage


\tableofcontents

\section{Introduction}
\setcounter{equation}{0}

The $c=1$ matrix model have been proved to be very powerful in
describing the two-dimensional string theory to all genus, both in
string perturbation theory and in a nonperturbative sense
\cite{GrossMiljkovic,BKZamo,GinsZ-J,Parisi}. The underlying rich
structure of the non-relativistic quantum mechanics of free
fermions (exhibited by the singlet sector) makes the theory
solvable to a very high degree and virtually renders any quantity
calculable to all orders in genus expansion. Recently the $c=1$
matrix model is realized as the effective dynamics of $D$-branes
in $c=1$ non-critical string theory which exhibits the duality
between $c=1$ matrix model and $2D$ quantum gravity coupled to
$c=1$ matter as an exact open/closed string duality
\cite{mgv,martinec,kms,mgtv,akk,TT,DKKMMS,sen-kyoto03,dd-singlet}. 
The quantum mechanics of $SU(N)$
invariant matrix variables in an inverted oscillator potential is
visualized as the quantum mechanics of open string tachyons
attached to $N$ unstable $D0$ branes that decays into ({\it i.e.}
dual to) Liouville theory coupled to $c=1$ matter describing $2D$
closed string theory together with its $D0$ branes.

The random surfaces mapped by the $c=1$ matrix model can also be
embedded in a circle of radius $R$ as a compactified Euclidean
theory or equivalently as a Minkowski signature string theory at
finite temperature, where the free fermion representation is not
sufficient due to the active role of the angular degrees of
freedom belonging to the nontrivial representation of $SU(N)$
\cite{GK1,GK2,BoulKaza}.
From matrix quantum mechanics analysis the states in the
nontrivial representation of $SU(N)$, the nonsinglet sector, are
understood to correspond to vortices on the world-sheet with wave
functions given by the Young Tableaux of $SU(N)$. The number of
boxes counts the vortex charge. Restricting the theory only to the
$SU(N)$ singlet sector gives rise to the continuum limit where the
effect of the world-sheet vortices are absent and the sum over the
random surfaces obey T-duality \cite{GK1}. Keeping the vortices
triggers Kosterlitz-Thouless phase transition at self-dual radius
that breaks the $R \to 1/R$ duality symmetry. Thus
the states in the nontrivial representation of $SU(N)$ are
important in understanding the dynamics of the world-sheet
vortices and are expected to give rise to interesting phenomena
like formation of $2D$ black hole \cite{KKK}.

An interesting solution of the two-dimensional string theory,
apart from the flat space background with linear dilaton, is the
two-dimensional black hole \cite{RabiBH,WadiaBH,WittenBH}.
Some of the early attempts to get matrix-model description of
two-dimensional black hole are described in
\cite{DMW1,Russo,Yoneya,Das,DMW2,JevYon,PolLH,Dhar}. It was partially
believed that the nonperturbative formulation of two-dimensional string
theory in terms of an integrable theory of noninteracting
nonrelativistic fermion representation of the matrix quantum
mechanics can deal issues like black hole evaporation and
gravitational collapse. But it turns out to be remarkably
difficult to find a matrix model description of the
two-dimensional black hole due to lack of clear
understanding of the target space physics of linear dilaton
background in the $c=1$ matrix model. The wave equation should
carry information about the black hole supplemented by the dilaton.
In free fermion representation,  the {\it
time-of-flight coordinate} $\tau$, related to the matrix model
coordinate $\lambda$ by $\lambda = \sqrt{2\mu\alpha'}\cosh\tau$,
cannot be identified with the Liouville coordinate $\phi$ although
this identification arises naturally from the collective field
theory approach \cite{DasJevicki,Pol-collective}. This 
is due to the fact that the
eigenvalue coordinate $\lambda$ has no obvious geometrical
interpretation on the discretized world-sheet. According to
\cite{Moore-macro,MooreSeiberg-macro},
the correct identification is through the loop operator
$W(l)$, which has a clear geometrical meaning on the world-sheet,
by an integral transform. Also it has been pointed out in \cite{Witten-cin} 
that exact solvable structure, especially the $W(\infty)$ symmetry 
of the $c=1$ model makes black hole hard to describe.   

One would hope to understand the situation better by working in a
more general representation including the world-sheet vortices.
However, the study of world-sheet vortices is hard due to lack of
solvable structure. In the spherical approximation, they could be
studied in dual matrix description considering discrete time
\cite{parisi,zaugg1,zaugg2}. In \cite{BoulKaza}, using twisted boundary
condition $\phi (2\pi R)=\Omega^\dagger \phi (0) \Omega$, the
partition function was studied in a given representation for the
standard matrix oscillator (with a stable quadratic potential).
The partition function in the presence of the adjoint
representation (a vortex-antivortex pair) in the double scaling
limit was then studied by analytically continuing to the upside
down oscillator . However, a direct analytical continuation is not
possible as the standard oscillator has larger symmetry than the
upside down one and does not have any information about the
cut-off provided by the interaction terms in the matrix potential.
Hence the analytical continuation had to be completed (in the
spirit of \cite{Moore}) by a suitable guess about the cut-off
dependence and was argued to work at least for the adjoint
representation.

Based on the above approach of connecting the world-sheet vortices
with the nonsinglet states of the matrix quantum mechanics, an
integrable system has been constructed which is an integrable Toda
chain hierarchy interpolating between the usual $c=1$ string and
the Sine-Liouville background \cite{KKK}\footnote{The description
is related by T-duality to the Toda integrable structure of the
$c=1$ string theory perturbed by purely tachyon source \cite{DMP}.}.
Using the duality conjecture by Fateev, Zamolodchikov and
Zamolodchikov (FZZ) \cite{FZZ-duality}, it was proposed to be a matrix
model (the KKK model) for two-dimensional black hole at
$R=3/2$ that relates the two-dimensional black hole
background ($SL(2,R)/U(1)$ coset CFT) to the condensation of
vortices (interpreted as the winding modes around the Euclidean
time). The effect of the winding modes on the world-sheet were
incorporated in the matrix quantum mechanics path integral by
integrating over the twist variables (the matrix holonomy factor
around the compactification circle) along with an appropriate
measure and the twisted partition function. The FZZ correspondence
allows one to consider the winding mode condensation from the
Sine-Liouville side to construct the appropriate matrix model for
the black hole background, avoiding dealing directly more complicated
black hole geometry. However, not considering the black hole background
directly has some disadvantages for the following reason. Although
FZZ conjecture has been tested by calculating various correlators,
it is not clear how to get information about the
black hole metric from the Sine-Liouville side. On the other hand,
considering the thermodynamics of two-dimensional string theory
above the temperature corresponding to the Berezinski-Kosterlitz-Thouless (BKT)
phase transition \cite{BKT1,BKT2,BKT3}, it has been proposed 
\cite{KKK} that the
nonsinglet states (or the vortices on the world-sheet) fills the Hagedorn
density of black holes $\rho (E) \sim E^{-s_1}e^{\beta_H E}$,
since the perturbative spectrum of two-dimensional string theory
has very few states, namely the massless tachyon. The Hagedorn
spectrum of states should then be obtained by a direct counting of
the nonsinglet states in the $c=1$ matrix model. From a
Hamiltonian point of view, to count the states in a given energy
interval one needs to diagonalize the Hamiltonians  for
representations corresponding to the large Young tableaux. This is
difficult due to Calogero type of interaction of the eigenvalues
with the $SU(N)$ spin structure. Also, there are other
important questions like understanding black hole phases at any
radius other than the fixed radius $R=3/2$ (to which the unstable
black hole can decay to), the possibility of observing Hawking
radiation and the like.

The main motivation of the present paper is to address above
questions from an explicit study of the nonsinglet sector and the
BKT phase transition directly from the matrix quantum mechanics
path integral with periodic boundary condition. Instead of working
in any particular representation let us have recourse to the
renormalization group approach that we developed in a recent paper
\cite{dd-singlet} for the $c=1$ matrix model on a circle of radius
$R$. There we described a detail analysis of how the two coupling
constants in the double scaling limit
with critical exponent flow with the change in length scale. The
motivation came from an earlier work of Br\'ezin and Zinn-Justin
\cite{BZ-J} on large $N$ RG analysis of $c=0$ matrix model. The
scheme reproduces the known results of the solvable sub-sector of
the matrix quantum mechanics, namely a non-trivial fixed point
with the correct string susceptibility exponent of the $c=1$
model, T-dulaity and the expected logarithmic scaling violation of
the free energy. Also it exhibits, qualitatively, the physically
interesting situations due to the effects of nonsinglet sector,
which can not be simplified because of the lack of the solvable
structure. For example, from the running of the prefactor of the
partition function, written in the renormalized couplings,
analogous to the running due to the wave function renormalization,
the free energy is observed to change sign near $R=1$ for small
value of the critical coupling. This is reminiscent of the
BKT transition at self-dual radius triggered by
the liberation of the world-sheet vortices. We would like to
understand the detail nature of the nontrivial fixed points of the
flow that describes the physics beyond this transition.

To capture the effect of vortices on the flows and the fixed
points more clearly and to introduce a new coupling that would act
like vortex fugacity, in this paper we analyze the behavior of the
following gauged matrix model with simple periodic boundary
condition and with an appropriate gauge breaking term

\ba &&\mathcal{Z}_{N}[g,\alpha,R]=\int_{\phi_{N}(2\pi
R)=\phi_{N}(0)}
\mathcal{D}^{(N)^2}A_{N}(t)~\mathcal{D}^{(N)^2}\phi_{N}(t)
\nonumber \\
&&\exp\Big[-(N)~\mbox{Tr} \int_0^{2\pi R} dt~
\Big\{\frac{1}{2}(D\phi_{N}(t))^2 + \frac{1}{2}\phi_{N}^2(t)
-\frac{g}{3}\phi_{N}^3(t)+\frac{A_{N}^2}{\alpha}\Big\}\Big]
\,,
\nonumber \\
\label{AZ1N} \ea
where the covariant derivative $D$ is defined with respect to the
pure gauge  $A(t) = \Omega(t)^\dagger\dot\Omega(t)$ by
$D\phi = \partial_t \phi + [A,\phi]$. Unlike taking the course of integration
over the twist fields with a proper measure incorporating winding modes around
the Euclidean time in the path integral (as considered in {\cite{KKK}}),
here the integration over the gauge field
$A(t)=\Omega^\dagger \dot \Omega (t)$ with an appropriate measure
provided by the gauge breaking term inserts world-sheet vortices
in the partition function where $\alpha$ acts as the vortex
fugacity. This can happen through the insertion of operator of the
form $\exp (- \alpha J^2)$ that
counts vortex number, where $J^2 \sim n N$.
Without the gauge breaking term the system is projected
to the singlet sector. We observe that one class of nontrivial
fixed points of the flow give rise to a pair of $c=1$ fixed points
at large $R$ with one unstable direction. As $R$ is decreased the
flow passes through Kosterlitz-Thouless phase transition at
$R=1.03$, where the operators coupled to the vortex fugacity become
relevant indicating the liberation of the world-sheet vortices as
expected at the self dual radius $R=1$ \cite{GK2}. Between $0.67 \le R \le
0.7$ the pair of fixed points become purely repulsive fixed points
of large coupling and exhibit negative specific heat and one loop
correction to the Hagedorn density of states very similar to those
exhibited by an unstable Euclidean black hole in flat space time.
The change of entropy exhibits a discontinuity at $R=0.73$, little
above the BKT temperature, indicating the Hagedorn transition to be
first order. As $R$ is further decreased the flow ends up into a
pair of $c=0$ fixed points. The other two classes of fixed
points also show black hole like behavior beyond the BKT phase
transition. This indicates he existence of other black hole phases
at other radii of compactification. Also a running of the
compactification radius with the scale, thought as dilaton, suggests
a deformation of the target space geometry that might be crucial
to visualize those black holes. We observe cyclic flow below the BKT
temperature presumably due to resonance of high spin states indicating
stringy behavior \cite{LeClair}. In fact, beyond the phase transition, 
where different nontrivial fixed points exhibit black hole like behavior,
the cycles become very complicated. The phase structure probed by the RG
analysis thus remarkably captures the expected properties and
looks promising in understanding the much unexplored physics of
the nonsinglet sector. The dynamics in the neighborhood of the
black hole like fixed point needs to be studied in detail to
understand its explicit nature. In this paper we will motivate
such studies regarding the role of the nonsinglet sector from the
observations from our Renormalization group analysis. 

The plan of the paper is as follows. In section 2 we describe the
RG calculation for our model (\ref{AZ1N}). In section 3 we analyze
the flow and discuss the phase structure. We observe the BKT phase
transition near the self dual radius (or a Hagedorn transition
little below the BKT radius) and existence of a black hole like
fixed point past the transition. In section 4 we analyze the
thermodynamics of the black hole like fixed point and comment on
the sense in which it resembles the thermodynamics of an unstable
(thermal) Euclidean black hole. In section 5 we conclude with
discussion and open questions.


\section{The Large $N$ RG Calculation}
\setcounter{equation}{0}

In this section we will summarize the renormalization group
calculation for the $c=1$ gauged matrix model on a circle, with a
gauge breaking term appropriate for capturing the nonsinglet physics.
The details of the method for the ungauged model with
periodic boundary condition is studied in \cite{dd-singlet}.

Let us consider the partition function for the $N+1$
dimensional matrix variables
\ba &&\mathcal{Z}_{N+1}[g,\alpha,R]=\int_{\phi_{N+1}(2\pi
R)=\phi_{N+1}(0)}
\mathcal{D}^{(N+1)^2}A_{N+1}(t)~\mathcal{D}^{(N+1)^2}\phi_{N+1}(t)
\nonumber \\
&&\exp\Big[-(N+1)~\mbox{Tr} \int_0^{2\pi R} dt~
\Big\{\frac{1}{2}(D\phi_{N+1}(t))^2 + \frac{1}{2}\phi_{N+1}^2(t)
-\frac{g}{3}\phi_{N+1}^3(t)+\frac{A_{N+1}^2}{\alpha}\Big\}\Big]
\,.
\nonumber \\
\label{AZ1_{N+1}} \ea The covariant derivative $D$ and the gauge
field $A(t)$ are defined respectively as \be D\phi = \partial_t
\phi + [A,\phi]\,, ~~~~ A(t) =
\Omega(t)^\dagger\dot\Omega(t)\,,~~~~\Omega\in U(N)\,.
\label{defA} \ee Expanding the covariant derivative, the partition
function is rewritten as \ba
\mathcal{Z}_{N+1}[g,\alpha,R]&=&\int_{\phi_{N+1}(2\pi
R)=\phi_{N+1}(0)}
\mathcal{D}^{(N+1)^2}A_{N+1}(t)~\mathcal{D}^{(N+1)^2}\phi_{N+1}(t)
\nonumber \\
&&\exp\Big[-(N+1) \mbox{Tr} \int_0^{2\pi R} dt~
\Big\{\frac{1}{2}\dot{\phi}_{N+1}(t)^2 +
\frac{1}{2}\phi_{N+1}^2(t) -\frac{g}{3}\phi_{N+1}^3(t)
\nonumber \\
&&+A_{N+1}(t)~[\phi_{N+1}(t),\dot{\phi}_{N+1}(t)]
+\frac{1}{2}~[A_{N+1}(t),\phi_{N+1}(t)]^2
+\frac{A_{N+1}^2}{\alpha}\Big\}\Big] \,.
\nonumber \\
&& \label{expAZ1_{N+1}} \ea
The $A_{N+1}(t)[\phi_{N+1}(t),\dot\phi_{N+1}(t)]$ term above is
crucial to study the nonsinglets. Even though they are present,
the gauge invariance tries to project the system to the singlet sector
while the gauge breaking term prevents to do so. In \cite{GK1},
a finitely large radius representation of singlet sector was obtained
by throwing this particular term by hand as the nonsinglets are
confined at small temperature. In \cite{BoulKaza}, the partition function
for one vortex/anti-vortex pair, {\it i.e.} in the adjoint representation
was calculated by analytical continuation from the twisted partition
function of the standard harmonic oscillator to that of the upside down
oscillator. For $\alpha =0$, the gauge fields
are forced to vanish and the partition function reduces
to that of ungauged matrix quantum mechanics on circle.

Because of the gauge breaking term, the integration over all
possible configurations of $A(t)$ formally inserts (the gauge
invariant) operator $\exp \mbox{Tr}(-\alpha J^2)$ in the partition
function, \be \int dA~\exp\mbox{Tr}\Big(-\frac{A^2}{\alpha}+2iAJ\Big)
~\sim~ \exp (- \alpha J^2) ~\sim~ \exp (-N\alpha n)\,. \ee Here
$J^2$ is proportional to the quadratic Casimir invariant $C(n)
\approx Nn$. Characterizing the irreducible representations  in
terms of the number of the white boxes $n$ in the Young tableaux,
the quadratic Casimir only depends on $n$ to the leading order in
$N$. The reason for this behavior is that,
$\exp \mbox{Tr}(-\alpha J^2)$ acts on the  states
$\vert\mbox{Adj}\rangle$ in the adjoint representation (belonging
to the nonsinglet sector) of the MQM in the gauge invariant way,
 \be A\cdot J\vert
\mbox{Adj}\rangle^n = \alpha~ n \vert \mbox{Adj}\rangle^n\,. \ee
The parameter $\alpha$ behaves like
the fugacity of vortices. The operator
$\exp \mbox{Tr} (-\alpha J^2)$ therefore counts the vortex number.


\subsection {Integrating out a column and a row of the matrices:}

Now we decompose the $(N+1)\times (N+1)$ matrices into $N\times N$
blocks and $N$-vectors and scalars as follows

\be \phi_{N+1}(t) = \begin{pmatrix} \phi_N(t) &  v_N(t) \cr
v^*_N(t) & \epsilon \cr \end{pmatrix} \,, \ee and
 \be
A_{N+1}(t) = \begin{pmatrix} A_N(t) &  a_N(t) \cr a^*_N(t) & \eta
\cr \end{pmatrix} \,. \ee The scalars $\epsilon$ and $\eta$ are of
relative order $1/N$ and can be ignored in the double scaling
limit. Even though the matrix elements are not independent ($\phi$
being hermitian and $A$ being antihermitian) one can always
choose such decomposition for each value of $N$ which
does not prevent the action
to be written in the same form
in terms of the covariant derivatives.
The resulting partition function can be written as

\ba \mathcal{Z}_N[g,\alpha,R]&=&\int_{\phi_N(2\pi R)=\phi_N(0)}
\mathcal{D}^{N^2}A_N(t)~\mathcal{D}^{N^2}\phi_N(t)
\nonumber \\
&&\exp\Big[-(N+1)~\mbox{Tr} \int_0^{2\pi R} dt~
\Big\{\frac{1}{2}\dot{\phi}_N(t)^2 + \frac{1}{2}\phi_N^2(t)
-\frac{g}{3}\phi_N^3(t)
\nonumber \\
&&+A_N(t)~[\phi_N(t),\dot{\phi}_N(t)]
+\frac{1}{2}~[A_N(t),\phi_N(t)]^2 +\frac{A_N^2}{\alpha}\Big\}\Big]
\,.
\nonumber \\
&&\times\int_{v,v^*(2\pi
R)=v,v^*(0)}\mathcal{D}^Na(t)\mathcal{D}^Na^*(t)
\mathcal{D}^Nv(t)\mathcal{D}^Nv^*(t)
\nonumber \\
&&\exp\Big[-(N+1)\int_0^{2\pi R}dt~\Big\{\dot v^*\dot v + \dot v^*
A_N v -v^* A_N \dot v + v^*(A_N^2+1-g\phi_N)v
\nonumber \\
&&+a^*\Big(\phi_N^2+\frac{1}{\alpha}\Big)a+v^*(\dot\phi_N+2A_N\phi_N-\phi_NA_N)a-\dot
v^*\phi_N a
\nonumber \\
&&+ a^*(-\dot\phi_N+2\phi_NA_N-A_N\phi_N)v+a^*\phi_N\dot v
\nonumber \\
&&+a^*va^*v+v^*av^*a-v^*va^*a-a^*vv^*a \Big\}\Big]\,.
\label{expAZ1_N} \ea Rescaling the vectors $v(t) \rightarrow
\frac{v(t)}{\sqrt{2\pi R(N+1)}}$ and $a(t) \rightarrow
\frac{a(t)}{\sqrt{2\pi R(N+1)}}$, the $v$ and $a$ dependent part
of the partition function turns out to be

\ba I[g,\alpha,\phi_N,A_N,R]&=&\frac{1}{(\sqrt{2\pi
R(N+1)})^{4N}}\int_{v,v^*(2\pi R)=v,v^*(0)}
\mathcal{D}^Na(t)\mathcal{D}^Na^*(t)
\mathcal{D}^Nv(t)\mathcal{D}^Nv^*(t)
\nonumber \\
&&\exp\Big[-\int_0^{2\pi R}\frac{dt}{2\pi R}~\Big\{\dot v^*\dot v
+ \dot v^* A_N v -v^* A_N \dot v + v^*(A_N^2+1-g\phi_N)v
\nonumber \\
&&+a^*\Big(\phi_N^2+\frac{1}{\alpha}\Big)a+v^*(\dot\phi_N+2A_N\phi_N-\phi_NA_N)a-\dot
v^*\phi_N a
\nonumber \\
&&+ a^*(-\dot\phi_N+2\phi_NA_N-A_N\phi_N)v+a^*\phi_N\dot v
\nonumber \\
&&+\frac{1}{2\pi R(N+1)}(a^*va^*v+v^*av^*a-v^*va^*a-a^*vv^*a)
\Big\}\Big]\,. \ea

Using the following Fourier transformation

\be
\mathcal{O}(t)=\sum_{m=-\infty}^{\infty}\mathcal{O}_m~e^{i\frac{m}{R}t}\,,~~~
\mathcal{O}_m = \int_0^{2\pi R}\frac{dt}{2\pi
R}~e^{-i\frac{m}{R}t} \mathcal{O}(t)\,, \label{FourierTrfm} \ee
with
$$
\delta_{mn} = \int_0^{2\pi R}\frac{dt}{2\pi
R}~e^{i\frac{(n-m)}{R}t}\,,~~~ \delta(t-t') = \frac{1}{2\pi
R}\sum_{m=-\infty}^{\infty}e^{i\frac{m}{R}(t-t')} \,,
$$
the $(v,a)$-integration can be expressed as

\ba && I[g,\alpha,\phi_N,A_N,R] = \frac{1}{(\sqrt{2\pi
R(N+1)})^{4N}}\int \Big(\prod_m da_m^*da_m^*\Big)\Big(\prod_m
dv_n^*dv_n^*\Big)
\nonumber \\
&&\exp\Big\{-\sum_m\Big(v_m^*\Big(\frac{m^2}{R^2}+1\Big)v_m+\frac{a_m^*a_m}
{\alpha}\Big)\Big\}
\nonumber \\
&&\exp\Big\{-\sum_{m,l}v_m^*\Big(-i\frac{m}{R}A_{m-l}-i\frac{l}{R}A_{m-l}
+\sum_k A_{m-l-k}A_k - g\phi_{m-l}\Big)v_l\Big\}
\nonumber \\
&&\exp\Big\{-\sum_{m,l} a_m^*(\sum_k\phi_{m-l-k}\phi_k)a_l\Big\}
\nonumber \\
&&\exp\Big\{-\sum_{m,l}v_m^*\Big(i\frac{(m-l)}{R}\phi_{m-l}+i\frac{m}{R}
\phi_{m-l}+2\sum_k A_{m-l-k}\phi_k - \sum_k \phi_{m-l-k}
A_k\Big)a_l\Big\}
\nonumber \\
&&\exp\Big\{-\sum_{m,l}a_m^*\Big(i\frac{(m-l)}{R}\phi_{m-l}+i\frac{m}{R}
\phi_{m-l}+2\sum_k \phi_{m-l-k}A_k - \sum_k A_{m-l-k}
\phi_k\Big)v_l\Big\}\,,
\nonumber \\
&& \ea where we have neglected the $O(1/N)$ terms. The above integration,
as an extra piece to the matrix integral, contains the vectors $v(t)$ (or quarks)
generating boundaries on the Feynman diagrams. In fact in vanishing $\alpha$
limit, this is essentially equivalent to the model considered in
\cite{Yang,Minahan,dd-singlet}. For generic value of $\alpha$, the above
integral should insert nonsinglet boundaries on the world-sheet that arise
after the BKT phase transition.


\subsection {One loop Feynman diagrams}

\begin{figure}[htb]
\centering
\includegraphics[width=450pt]{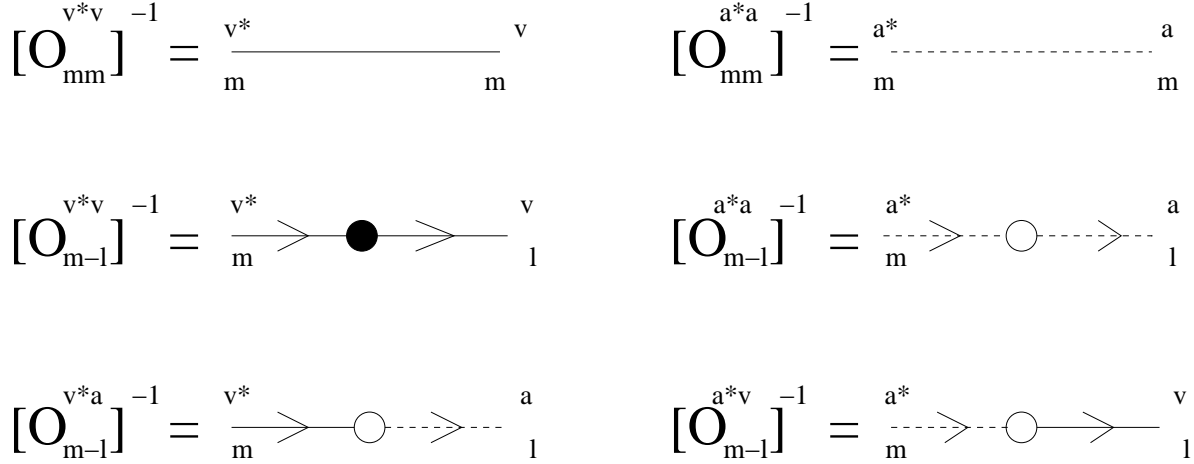} 
\caption{
The propagators and vertices.
} \label{prop}
\end{figure}

In order to carry out the $v$ and $a$ integration
diagrammatically, let us now define the following operators

\ba \mc{O}_{mn}^{v^*v} &=&
\Big(\frac{mn}{R^2}+1\Big)\delta_{mn}\,,~~~
 \mc{O}_{mn}^{a^*a} (\alpha) = \frac{\delta_{mn}}{\alpha}\,,
 \nonumber \\
 \mc{O}_{m-l}^{v^*v}(g,\phi,A)
 &=&\Big(-i\frac{(m+l)}{R}A_{m-l}+\sum_kA_{m-l-k}A_k-g\phi_{m-l}\Big)\,,
 \nonumber \\
 \mc{O}_{m-l}^{a^*a}(\phi)&=&\Big(\sum_k\phi_{m-l-k}\phi_k\Big)\,,
 \nonumber \\
 \mc{O}_{m-l}^{v^*a}(\phi,A)&=&\Big(i\frac{(2m-l)}{R}\phi_{m-l}
 +2\sum_kA_{m-l-k}\phi_k
 -\sum_k\phi_kA_{m-l-k}\Big)\,,
 \nonumber \\
 \mc{O}_{m-l}^{a^*v}(\phi,A)&=&\Big(i\frac{(2l-m)}{R}\phi_{m-l}
 +2\sum_k\phi_kA_{m-l-k}
 -\sum_kA_{m-l-k}\phi_k\Big)\,.
\ea The inverse of these operators define various propagators and
vertices according to figure \ref{prop}.

\begin{figure}[htb]
\centering
\includegraphics[width=450pt]{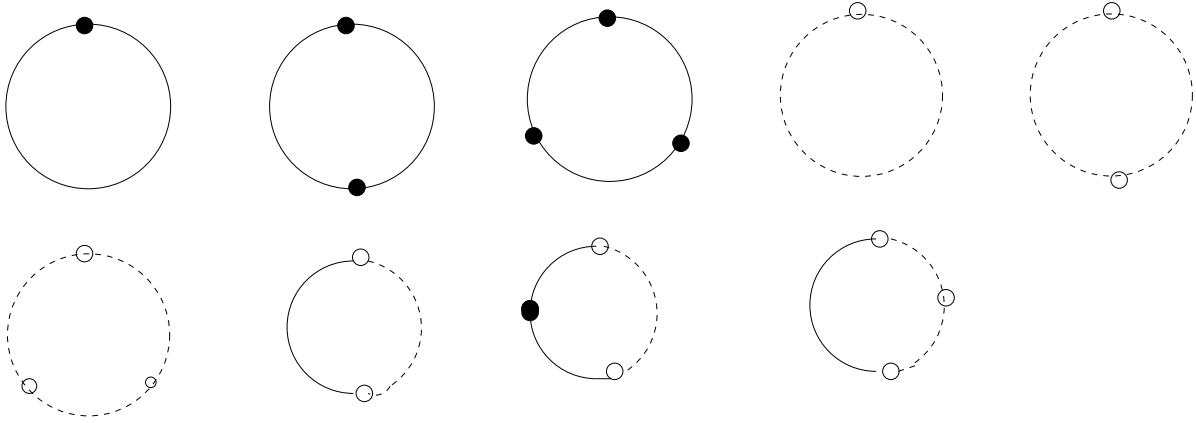} 
\caption{
The one loop
  diagrams contributing to $\Sigma$.
} \label{feynman}
\end{figure}

Hence the integral becomes

\ba I[g,\alpha,\phi_N,A_N,R,N]&=&\frac{1}{(2\pi
R(N+1))^{2N}}~\exp\Big[-\{\sum_{m_1,l_1}
(\mc{O}_{m_1-l_1}^{v^*v}(g,\phi,A))^{-1}
\nonumber \\
&&+\sum_{m_2,l_2}(\mc{O}_{m_2-l_2}^{a^*a}(\phi))^{-1}
+\sum_{m_3,l_3}(\mc{O}_{m_3-l_3}^{a^*v}(\phi,A))^{-1}
\nonumber \\
&&+\sum_{m_4,l_4}(\mc{O}_{m_4-l_4}^{v^*a}(\phi,A))^{-1}
\}\Big]~I_0(\alpha,R,N)\,, \label{I1} \ea where the gaussian part
is as follows

\be I_0[\alpha,R,N]=\int \Big(\prod_i
da_i^*da_i\Big)\Big(\prod_jdv_j^*dv_j\Big)
~\exp\big[-\sum_m(v_m^*\mc{O}_{mm}^{v^*v}v_m+a_m^*\mc{O}_{mm}^{a^*a}a_m)\big]\,.
\ee Performing the gaussian integration, we get

\be I_0[\alpha,R,N]=(2\pi
R)^{2N}\pi^{2N}\exp\Big[-\mbox{Tr}\Big\{\sum_m\ln
\Big(\frac{m^2}{R^2}+1\Big){\bf 1}\Big\}\Big]
\exp\Big[-\mbox{Tr}\Big\{\ln \Big(\frac{1}{\alpha}\Big){\bf
1}\Big\} \Big]\,. \ee Inserting this into (\ref{I1}), the
$(v,a)$-integration becomes

\be
I[g,\alpha,\phi_N,A_N,R,N]=\mc{C}(\alpha,R,N)~\Sigma[g,\alpha,\phi_N,A_N,R,N]\,,
\label{I2} \ee where

\be \mc{C}(\alpha,R,N)=\bigg[\frac{\pi^3R\alpha}{(N+1)^2\sinh\pi
R}\bigg]^N ~\frac{1}{\prod_m(m/R)^{2N}}\,. \ee In (\ref{I2}),
$\Sigma[g,\alpha,\phi_N,A_N,R,N]$ represents sum of one loop
Feynman diagrams as shown in figure \ref{feynman}.

\subsection{ Evaluation of the diagrams}

We evaluate the diagrams to express  $\Sigma
[g,\alpha,\phi_N,A_N,R,N]$ as a series expansion in the couplings,
$g$, $\alpha$, and the Fourier modes of the fields
$\phi_{m}$, $A_{m}$. The full series expansion is  given in the
Appendix A.  We will now discuss how to express this series into
an expansion in $g$, $\alpha$, $\Phi(t)$ and $A(t)$ by suitable
inverse Fourier transform.

In order to evaluate the one loop correction to the effective
action, we inverse transform the Fourier modes according to the
rule (\ref{FourierTrfm}) and sum-up the set of infinite series
using the formulae discussed below.

\ba \sum_{m=-\infty}^{\infty}
\frac{\exp[i(m/R)t]}{\frac{m^2}{R^2}+a^2} &=& \frac{\pi
R}{a}\frac{\cosh(\pi aR-at)}{\sinh\pi aR} \,,~~~0\le t\le 2\pi R
\,,
\nonumber \\
\sum_{m=-\infty}^{\infty}\frac{m/R~\exp[i(m/R)t]}{\frac{m^2}{R^2}+a^2}
&=& \frac{i\pi R~\sinh(\pi a R - a t)}{\sinh \pi a R}\,,~~~ 0< t<
2\pi R \,,
\nonumber \\
&& i\pi R \,,~~~~~~~~~~~~~~~~~~~~~~~~~~t \to 0 \,,
\nonumber \\
&& -i\pi R \,,~~~~~~~~~~~~~~~~~~~~~~~t \to 2\pi R \,,
\nonumber \\
&& 0 \,,~~~~~~~~~~~~~~~~~~~~~~~~~~t=0=2\pi R\,. \ea Because of the
discontinuity in the last series shown above, all other series,
which look like higher derivatives of the above functions,  has
delta function or derivative of delta function like behavior. This
behavior actually dominate over the regular part of those
functions and hence dominate the contribution when the functions
are integrated over :

\ba
\sum_{m=-\infty}^{\infty}\frac{m^2/R^2~\exp[i(m/R)t]}{\frac{m^2}{R^2}+a^2}
& \sim & 2\pi R\delta(t) \,, ~~~0\le t\le 2\pi R\,,
\nonumber \\
\sum_{m=-\infty}^\infty
\frac{m^3/R^3~\exp[i(m/R)t]}{\frac{m^2}{R^2}+a^2} & \sim & -2\pi
iR\delta'(t) \,, ~~~0<t<2\pi R\,,
\nonumber \\
\sum_{m=-\infty}^\infty
\frac{m^4/R^4~\exp[i(m/R)t]}{\frac{m^2}{R^2}+a^2} & \sim & -2\pi
R\delta''(t) \,, ~~~0\le t\le 2\pi R\,. \ea One way to see the
above behavior is as follows:

\ba
\sum_{m=-\infty}^{\infty}\frac{m^2/R^2~\exp[i(m/R)t]}{\frac{m^2}{R^2}+a^2}
& \sim & \sum_{m=-\infty}^\infty \exp[i(m/R)t] = 2\pi R\delta(t)
\,,~~~0\le t\le 2\pi R\,,
\nonumber \\
\sum_{m=-\infty}^{\infty}\frac{m^2/R^2}{\frac{m^2}{R^2}+a^2}
&\sim& \sum_{m=-\infty}^{\infty} 1 = 2\pi R\delta(0) \,. \ea Also

\ba \sum_{m=-\infty}^\infty
\frac{m^3/R^3~\exp[i(m/R)t]}{\frac{m^2}{R^2}+a^2} & \sim &
\sum_{m=-\infty}^\infty\frac{m}{R}~\exp[i(m/R)t] = -2\pi
iR\delta'(t) \,, ~~~0<t<2\pi R\,,
\nonumber \\
\sum_{m=-\infty}^\infty\frac{m^3/R^3}{\frac{m^2}{R^2}+a^2} & \sim
& \sum_{m=-\infty}^\infty\frac{m}{R} = 0 \,,
\nonumber \\
\sum_{m=-\infty}^\infty
\frac{m^4/R^4~\exp[i(m/R)t]}{\frac{m^2}{R^2}+a^2} & \sim &
\sum_{m=-\infty}^{\infty}\frac{m^2}{R^2}~\exp[i(m/R)t]=-2\pi
R\delta''(t) \,, ~~~0\le t\le 2\pi R \,.
\nonumber \\
\sum_{m=-\infty}^\infty \frac{m^4/R^4}{\frac{m^2}{R^2}+a^2} & \sim
& \sum_{m=-\infty}^{\infty}\frac{m^2}{R^2} = -2\pi R\delta''(0)
\,. \ea
Note that while integrating over these functions, we have
to take into account their periodic nature and the intervals over
which they are defined. We also have to deal with subtleties when
these intervals are strictly inequalities.

These sums give the
corrections to the coefficients of the various terms in the
action, after $\Sigma[g,\alpha,\phi_N,A_N,R,N]$ is exponentiated
and log-expanded around $\phi=0$ (small field approximation).
Since after doing the inverse Fourier transform,
the various terms containing $\phi(t)$ and $A(t)$ has nonlocal integrals
over several one dimensional dummy time variables, we breakup the
variables into center of mass and relative coordinates. Then we
expand the functions about the center of mass coordinates,
assuming the relative coordinates to be small enough, and consider
integration over the relative coordinates.

Now taking into account all these considerations, after evaluating
the integrations over relative time variables (see Appendix B),
the expression for $\Sigma[g,\alpha,\phi_N,A_N,R,N]$ becomes

\ba &&\Sigma[g,\alpha,\phi_N,A_N,R,N] = 1+ F_{g1}(R)~g
\int_0^{2\pi R} dt~\mbox{Tr}\phi(t)
\nonumber \\
&&+\{F_{g2}(R)~g^2+F_{\alpha2}(R)~\alpha\}\int_0^{2\pi R}dt
~\mbox{Tr}\frac{1}{2}\phi^2 (t)
+\{\hat{F}_{g2}(R)~g^2+\hat{F}_{\alpha
2}(R)~\alpha\}\int_0^{2\pi R}dt ~\mbox{Tr}\frac{1}{2}\dot\phi^2(t)
\nonumber\\
&&+G_2(R)\int_0^{2\pi R}dt~\mbox{Tr}A^2(t)
+\{G_{g3}(R)~g^2 +G_{\alpha3}(R)\alpha\}\int_0^{2\pi
R}dt~\mbox{Tr}A(t) \phi(t) \dot\phi(t)
\nonumber \\
&&+\{G'_{g3}(R)~g^2 +G'_{\alpha3}(R)\alpha\}\int_0^{2\pi
R}dt~\mbox{Tr}A(t) \dot\phi(t) \phi(t)
+\{F_{g3}(R)~g^3
\nonumber \\
&&+F_{\alpha g3}(R)~\alpha g\}\int_0^{2\pi
R}dt~\mbox{Tr}\frac{1}{3}\phi^3(t)
+\{G_{\alpha 4}(R)~\alpha+G_{g4}(R)~g^2\}\int_0^{2\pi
R}dt~\mbox{Tr}A(t)\phi(t)A(t)\phi(t)
\nonumber \\
&&+\{G'_{\alpha 4}(R)~\alpha+G'_{g4}(R)~g^2\}\int_0^{2\pi
R}dt~\mbox{Tr}A^2(t)\phi^2(t)\,.
\nonumber \\
\ea The functions $F(R)$s and $G(R)$s are
defined as follows

\ba
 F_{g1}(R)&=& \coth \pi R\,,~~~~
F_{g2}(R)= \frac{\pi R}{2}(1+\coth^2\pi R)
+\frac{1}{2}\coth\pi R\cosh 2\pi R
\nonumber\\
 F_{\alpha 2}(R) &=& -1/(\pi R)+8 - 4\coth \pi
 R\,,~~~~\hat F_{\alpha 2}(R)= -4 \coth \pi R\,,
 \nonumber\\
 \hat F_{g2}(R) &=&
 \frac{1}{\sinh^2 \pi R}\big(\frac{1}{64}(1+8\pi^2 R^2)
\sinh 4\pi R-\frac{\pi^3 R^3}{6}\cosh 2\pi R
-\frac{\pi R}{16}\cosh 4\pi R\big)\,,
 \nonumber\\
G_{2}(R)&=& -3 \coth \pi R+\frac{\pi R}{2 \sinh^2\pi R}
-\frac{\pi R}{2}(1+\coth^2\pi R)\,,
 \nonumber\\
 F_{g3}(R)&=& \frac{\pi R}{64\sinh^3\pi R(\cosh 2\pi R+\cosh 4\pi R)}
[4\pi R(3\cosh\pi R+2\cosh 3\pi R+2\cosh 5\pi R
\nonumber \\
&&+\cosh 7\pi R)
+\sinh\pi R +\sinh 5\pi R +2\sinh 7\pi R+\sinh 9\pi R]\,,
\nonumber\\
F_{\alpha g3}(R)&=& \frac{39\pi R}{4}(1+\coth^2\pi R)
-\frac{33}{4}\coth\pi R \,,
\nonumber\\
G_{g3}(R)&=& G_{g3}'(R)
= \frac{1}{12}(\pi R\cosh 2\pi R-\frac{1}{2}\sinh 2\pi R)\,,
\nonumber \\
G_{g4}(R)&=& -\frac{1}{8\sinh^2\pi R}\big(\pi^2 R^2\cosh 3\pi R
+\frac{\pi R}{8}\sinh 5\pi R+\frac{\pi R}{4}\sinh 3\pi R
-\frac{3}{8}\sinh \pi R\big)\,,
\nonumber \\
G_{\alpha 3}(R)&=& G_{\alpha 3}'(R)=-\frac{2}{3} \coth \pi R 
+ \frac{1}{\sinh^2\pi R}(\frac{\pi R}{6}\cosh\pi R
+\frac{1}{12}\sinh\pi R)\,,
\nonumber \\
G_{\alpha 4}(R) &=& -2 \coth \pi R\,, 
~~~~G_{\alpha 4}'(R) = \frac{5}{2}\coth\pi R\,,~~~~G_{g4}'(R)=0\,.
\label{defhyperbolic} \ea


\subsection{Elimination of the tadpole term}

 The term proportional to $\int_0^{2\pi R}dt~\phi(t)$ is unwanted.
To remove this term, we change the background by $\phi(t) \to
\phi(t)+f$, and set the net coefficient of the linear term to zero.
This fixes the value of $f$ to be

\ba f_{\pm} &=& \frac{1}{2}\Big(g+\frac{g}{N}+\frac{\alpha gF_{\alpha
g3}+g^3F_{g3}}{N}\Big)^{-1}\Big[\Big(1+\frac{1}{N}-\frac{\alpha
F_{\alpha 2}+g^2F_{g2}}{N}\Big)
\nonumber\\
&&\pm \Big\{\Big(1+\frac{1}{N}-\frac{\alpha F_{\alpha
2}+g^2F_{g2}}{N}\Big)^2
-\frac{4gF_{g1}}{N}\Big(g+\frac{g}{N}+\frac{\alpha gF_{\alpha g
3}+g^3F_{g3}}{N}\Big)\Big\}^{\frac{1}{2}}\Big] \ea As $f_{+}\sim
O(1)$ and $f_{-} \sim O(1/N)$, we choose the smaller shift \be
f_{-} \sim g F_{g1}/N \ee to suppress the contribution from the
higher order terms. Note that the coefficient of $\phi^n(t)$ term
would contribute a term proportional to $f$ in the coefficient of
$\phi^{n-1}(t)$ term. Thus the coefficient of $\phi^3(t)$ in the
coupling $g_3$ would have an $O(1/N)$ contribution in the coupling
$g_4$ from the $\phi^4(t)$ term, if we had turned it on. Also the
contribution from the $[A(t),\phi(t)]^2$ term in the coefficient
of the $A^2(t)$ term can be ignored as it is of $O(1/N^2)$. After
accommodating all such changes, the expression for $Z_{N+1}$
becomes

\ba &&Z_{N+1}=\mc{C}(\alpha,R,N)~\exp[2\pi
RN^2\mathscr{F}(\alpha,g,R,N)]
\int\mc{D\phi}\mc{DA}~\exp\Big[-N\mbox{Tr}\int_0^{2\pi R}dt~\big\{(1+1/N
\nonumber \\
&&-(\alpha\hat{F}_{\alpha
2}+g^2\hat{F}_{g2})/N)\dot\phi^2(t)/2+(1+1/N-(\alpha
F_{\alpha 2}+g^2(F_{g2}+F_{g1}))/N)\phi^2(t)/2
\nonumber \\
&&-(g+g/N+(\alpha gF_{\alpha g3}+g^3F_{g3})/N)\phi^3(t)/3
+(1/\alpha+1/(\alpha N)-G_2/N)A^2(t)
\nonumber\\
&&+(1+1/N-(\alpha G_{\alpha 3}+g^2G_{g3})/N)A(t)\phi(t)\dot\phi(t)
\nonumber\\
&&-(1+1/N+(\alpha G'_{\alpha 3}+g^2G'_{g3})/N)A(t)\dot\phi(t)\phi(t)
\nonumber \\
&&+(1+1/N-(\alpha G_{\alpha
4}+g^2G_{g4})/N)A(t)\phi(t)A(t)\phi(t)
\nonumber\\
&&-(1+1/N+(\alpha G'_{\alpha4}
+g^2G'_{g4})/N)A^2(t)\phi^2(t)\big\}\Big]\,.\label{renorm-action1}
\ea Here the expression for
$\mathscr{F}(\alpha,g,R,N)$ is given by

\ba \mathscr{F}(\alpha,g,R,N) &=&
\frac{gF_{g1}}{N}f-\Big(1+\frac{1}{N}-\frac{\alpha F_{\alpha 2}
+g^2F_{g2}}{N}\Big)\frac{f^2}{2}+\Big(g +\frac{g+\alpha gF_{\alpha
g3}+g^3F_3}{N}\Big)\frac{f^3}{3}
\nonumber\\
&&\sim g^2 F_{g1}^2/N^2+O(1/N^3)\,.\ea

\subsection{ Rescaling of the fields and the variables}

We will now rescale the fields and the variables
(time $t$ and the conjugate momentum $1/R$) to restore
the original cut-off (in the Wilsonian sense):

\ba
&& t \to t'(1+h~dl)\,,~~~~~R \to R'(1+h~dl)\,,
\nonumber\\
&&\phi(t) \to \rho\phi'(t')\,,
\nonumber\\
&&A(t) \to (1-h~dl)~\eta A'(t')\,, \label{rescaling} \ea where,
\be dl=1/N\,,~~~~h = h(R) + \sum_{i,j} c_{ij}~g^i\alpha^jh_{ij}(R)
\,. \ee The exact functional form of $h$ can be guessed from the
behavior of the Feynman diagrams. We will see that $h(R)$ is the
scaling dimension of the operator coupled with  mass parameter,
the coefficient of the $\phi(t)^2/2$ term ($1/\alpha'$, we have
set $\alpha'=1$ here), and is appearing in the universal term of
the beta function equation of the  mass parameter.  It is
interesting that we will also see $h(R)$ to  appear in combination
with other numbers as the scaling dimensions of the operators
coupled with the cosmological constant $g$, and the fugacity
$\alpha$. Therefore, being in the universal term of the beta
function equations of the couplings $g$ and $\alpha$, $h(R)$
determines the radius at which the corresponding operators become
relevant and could trigger a phase transition. The latter will
happen if  there is any discontinuous change in the free energy
like the flipping of sign. In  the world-sheet picture one expects
a phase transition at the self dual radius $R=1$ due to the
liberation of the world-sheet vortices \cite{GK1}. The world-sheet
free energy changes sign at that radius due to a contest between
the entropy of the liberated vortices and the energy of the
system. We will come back to this in the discussion of the beta
function equation of the fugacity parameter $\alpha$ and will see
that its universal term do reflect such a transition. Also through
the rescaling relation of $R$ (\ref{rescaling}), $h(R)$ determines
the change of the radius with the scale and hence will play a role
in discussing the thermal properties of the fixed points of the RG
flow.

Now in doing the rescaling of time, as shown in (\ref{rescaling}),
the field $A(t)$ automatically peaks up a factor of $(1-h~dl)$
because of the definition (\ref{defA}). Since there is no other
constraint on $A(t)$, we are here free to rescale it by an
arbitray factor $\eta$.  The rescaled action therefore looks like,
\ba S'=N \mbox{Tr} \int_{0}^{2 \pi R'} dt' \Big[ (1-h dl) \rho^2
\Big(\frac{\hat c_{2}}{2} \dot \Phi'(t')^2 +\hat c_{3} ~\eta
A'(t')\Phi'(t')\dot \Phi'(t')- \hat c'_{3} ~\eta
A'(t')\dot\Phi'(t')\Phi'(t')
\nonumber\\
+\hat c_{4} ~\eta^2 A'(t') \Phi'(t') A'(t') \Phi'(t')-\hat
c'_{4}~\eta^2 A^{'2}(t') \Phi^{'2}(t') \Big)
\nonumber\\
(1+h dl) \rho^2 \frac{ c_{2}}{2}\Phi'^2(t')-\frac{g}{3}(1+h dl)
\rho^3 c_{3} \Phi'^3(t') +(1-h dl) c_{4}~\frac{\eta^2}{\alpha}
A'(t')^2 \Big]\,.
\nonumber\\
\ea
The coefficients $\hat c_{i}$ and $c_{i}, ~i=2,3,4$ can be read
off by comparing with that of the renormalized action
in (\ref{renorm-action1}).
Setting the coefficient of the kinetic term
$\dot\phi^2/2$ to one, $\rho$ can be fixed as

\be \rho = 1+\frac{1}{2}[h-1+g^2\hat F_{g2}+\alpha\hat F_{\alpha
2}]dl + O(dl^2)\,. \ee
Similarly setting the coefficients of $A(t)[\phi(t),\dot\phi(t)]$ and that
of the $[A(t),\phi(t)]^2/2$
term respectively to one, we have
\be
\frac{\hat c_{3}}{\hat c_{2}} \eta=\frac{\hat c'_{3}}{\hat c_{2}} \eta=1\,,~~~
\frac{\hat c_{4}}{\hat c_{2}} \eta^2=\frac{\hat c'_{4}}{\hat c_{2}} \eta^2=1\,.
\ee
In other words, this fixes $\eta$ to

\be
\eta=\frac{\hat c_{2}}{\hat c_{3}} =1+[(G_{g3}
-\hat F_{g2}) g^2+(G_{\alpha 3}-\hat F_{\alpha_2})\alpha] dl+O(dl^2)\,,
\ee
along with the constraints
\be
\hat c_{3}=\hat c'_{3}\,,~~~\hat c_{4}=\hat c'_{4}\,,
~~~\hat c_{3}^2=\hat c_{2} ~\hat c_{4}\,.
\ee
Accordingly the coefficients of $\phi^2(t)/2$, $g\phi^3(t)/3$,
and $(1/\alpha)A^2(t)$ respectively are modified as

\be c_{2}+\delta c_{2}=(1+2 h~dl) \frac{c_{2}}{\hat c_{2}}
=1+[2h+(\hat F_{g2}-F_{g2}-F_{g1}) g^2
+(\hat F_{\alpha 2}-F_{g2})\alpha] dl + O(dl^2) \,,
\ee

\ba c_{3}+\delta c_{3}= (1+5h/2~dl)\frac{c_{3}}{\hat c_{2}^{3/2}}
=1+[5h/2-1/2+(3\hat F_{g2}/2+F_{g3}) g^2+(3\hat F_{\alpha
2}/2+F_{\alpha g3})\alpha] dl
\nonumber\\+ O(dl^2) \,,
\nonumber\\
\ea
and,
\be
c_{4}+\delta c_{4}=(1+h~dl)\frac{\hat c_{2}^2~c_{4}}{\hat c_{3}^2}
=1+[(1-h)+2(G_{g3}-\hat F_{g2})g^2+(2G_{\alpha3}
-2\hat F_{\alpha 2}-G_{2})\alpha]dl
+ O(dl^2)\,.\ee
Setting the coefficient of $\phi(t)^2/2$ to one
(this implies keeping the mass parameter
at the fixed point with unit magnitude)
results in an extra constraint
\be
\delta c_{2}=0\,.
\ee


\subsection{ Beta function equations}

The effective action is of the same form as the bare one, but with
renormalized strength of the coupling and the fugacity. The resulting
partition function is given by

\ba \mathcal{Z}_{N+1}[g',\alpha',R']&=&\lambda'^{N^2}\int_{\phi'_{N+1}(2\pi
R')=\phi'_{N+1}(0)}
\mathcal{D}^{(N+1)^2}A'_{N+1}(t')~\mathcal{D}^{(N+1)^2}\phi'_{N+1}(t')
\nonumber \\ &&\exp\Big[-N~\mbox{Tr} \int_0^{2\pi R'} dt'~
\Big\{\frac{1}{2}(D\phi'_{N+1}(t'))^2
+\frac{1}{2}\phi'^2_{N+1}(t') -\frac{g'}{3}\phi'^3_{N+1}(t')
\nonumber \\ &&+\frac{A'^2_{N+1}(t')}{\alpha'}\Big\}\Big] \,,
\nonumber \\
&& \label{A'Z1_{N+1}} \ea where
\be\lambda'^{N^2}=\mc{C}(R,N)\exp[-2\pi
RN^2\mathscr{F}(g,N,R)]\rho^{N^2}\,.\ee Neglecting the $O(dl^2)$
terms, the renormalized fugacities (couplings) and the vacuum energy
(the prefactor of the partition function) are expressed in
terms of the bare quantities as follows:

\ba g' &=& g+[(5h/2-1/2)g+(3\hat F_{g2}/2+F_{g3}) g^3+(3\hat F_{\alpha
2}/2+F_{\alpha g3})\alpha g] dl
\,, \nonumber \\
\frac{1}{\alpha'} &=&
\frac{1}{\alpha}+[(1-h)\frac{1}{\alpha}+2(G_{g3}-\hat
F_{g2})\frac{g^2}{\alpha}+(2G_{\alpha3}-2\hat F_{\alpha
2}-G_{2})]dl
\nonumber\\
\lambda'&=& 1+\ln \Big[\frac{\pi^3 R^2 \alpha}{\sinh \pi R }\Big] dl
+\frac{1}{2}\big[(h-1)+g^2 \hat F_{g2}+\alpha F_{\alpha 2}\big]dl\,. \ea Here, in
simplifying the part $\mc{C}(R,N)^{\frac{1}{N^2}}$ in the
expression for $\lambda'$, we have assumed that for any value of
$R$, \ba \mc{C}(R,N)^{\frac{1}{N^2}}&&=\exp \Big[\frac{1}{N} \ln
\Big(\frac{\pi^3 R^2 \alpha}{\sinh \pi R }\Big) \Big]
\Big[\frac{\pi^3R^{4n}}{ (N+1)(n!)^4}\Big]^{\frac{1}{N}}_{n \to
\infty,~N \to \infty}
\nonumber\\
&& \simeq 1+\frac{1}{N} \ln \Big(\frac{\pi^3 R^2 \alpha}{\sinh \pi R
M}\Big)+O(1/N^2)\,. \label{prefactor} \ea Also, the term
$\exp[-2\pi RN^2\mathscr{F}(g,N,R)]^{\frac{1}{N^2}}$ contributes only a factor
of $1$ as $\mathscr{F}(g,N,R)\sim O(dl^2)$.
Hence the resulting beta function equations are
expressed as

\ba \beta_g &=& \frac{dg}{dl} =
\big(\frac{5}{2}h-\frac{1}{2}\big)g +\big(\frac{3}{2}\hat
F_{\alpha 2}+F_{\alpha g3}\big)\alpha g
+\big(\frac{3}{2}\hat F_{g2}+F_{g3}\big)g^3\,, \nonumber \\
\beta_\alpha &=& \frac{d\alpha}{dl} = -(1-h)\alpha - (2G_{\alpha
3}-2\hat F_{\alpha 2}-G_2)\alpha^2-2g^2\alpha(G_{g3}-\hat
F_{g2})\,,
\nonumber\\
\beta_{\lambda}&=&\frac{d\lambda}{dl}=\ln \Big[\frac{\pi^3
R^2 \alpha}{\sinh \pi R }\Big] +\frac{1}{2}\big[(h-1)+g^2 \hat
F_{g2}+\alpha F_{\alpha 2}\big]\,.
\label{beta1}\ea
along with the constraints,
\ba
\alpha(G'_{\alpha 3}-G_{\alpha 3})+g^2(G'_{g3}-G_{g3})&=&0\,,
\nonumber\\
\alpha(G'_{\alpha 4}-G_{\alpha 4})+g^2(G'_{g4}-G_{g4})&=&0\,,
\nonumber\\
\alpha(\hat F_{\alpha2}+G_{\alpha 4}-2G_{\alpha3})+g^2(\hat F_{g2}+G_{g4}
-2G_{g3})&=&0\,,
\nonumber\\
2h+g^2(\hat F_{g2}-F_{g2}-F_{g1})
+\alpha(\hat F_{\alpha 2}-F_{\alpha 2})&=&0 \,.
\label{constraints}
\ea
The relation (\ref{rescaling})
indicates in some sense a running of the radius $R$  \be
\beta_R = \frac{d R}{dl}= -h R\,.\label{betaR}\ee
This suggests a deformation of the target space geometry
if one considers the scale to be dilaton.
In the above equations, all the
functions $F$s and $G$s are hyperbolic nonlinear functions of $R$,
as defined in (\ref{defhyperbolic}).

Now before going to the detail analysis of the fixed points let us
try to understand few things about the structure of the beta
function equations. The much of the structure depends on
understanding the quantity $h$. One can clearly see for
$g=0$ and  $\alpha=0$, the
gaussian model is never expected to flow and hence for such a
fixed point $h=0$~({\it i.e.} corresponding to the trivial
rescaling $t'=t$ and $R'=R$). The situation is different for a
non-vanishing $h$. Demanding the mass parameter $M^2$
(the coefficient
of the $\phi^2(t)/2$ term) to be set at
some fixed value (we will use 1 for simplicity) and not running,
one can easily determine some $h=h(R)$ for
nontrivial fixed points from the set of the beta function equations
and the set of constraints. This will be discussed in the next section.
This is extremely interesting as it could indicate a
phase transition at certain radius due to turning on the different operators
coupled to the relevant couplings.
Note that in (\ref{beta1}) the linear term in $\alpha$ in
$\beta_{\alpha}$ changes sign at $h=1$. This indicates that the
coupling corresponding to the vortex fugacity becomes relevant in the $h \ge
1$ region. In the next section, we will study this as an indication
of the liberation of the world-sheet vortices due to Kosterlitz-Thouless
type of phase transition undergone by the $c=1$ matrix quantum mechanics
with nonsinglet sector.

Keeping $M^2=1$ for simplicity is consistent with the value of the
mass parameter ($M^2=\frac{1}{\alpha'}$) one originally works with
in the matrix partition function to visualize the matrix path
integral as the generator of the discretized version of the
Polyakov path integral of $2D$ bosonic string. In recent
identification of the matrix quantum mechanics with the quantum
mechanics of open string tachyon on unstable $D0$-branes the mass
parameter $M^2=\frac{1}{\alpha'}$ is identified with the open
string tachyon mass. A framework of the flow
of a general $M^2$ ({\it i.e.} $h=0$~) could be interesting to discuss
the presence of the boundaries.


\section{Analysis of Flow Equations and Phase Transition}
\setcounter{equation}{0}


\subsection{The fixed points}
The fixed points of the flow are given by the simultaneous solution
of the beta function equations $\beta_g=\beta_{\alpha}=0$~. The Gaussian
fixed point $\Lambda^*_0$  is given by

\be
g^*=0\,,~~~~\alpha^*=0\,,~~~~h=0\,.
\label{lambda0}
\ee
The nontrivial fixed points of the flow $\Lambda^*_1(g^{*2} \ne 0, 0)$,
$\Lambda^*_2(0, \alpha^* \ne 0)$ and $\Lambda^*_3(g^{*2}\ne 0, \alpha^* \ne 0)$
are as follows,
\ba
&&\Lambda^*_1=\Big(2/(\hat F_{g2}-F_{g2}-F_{g1}-2F_{g3}-3G_{g4}),0; R_1\Big)\,,
\nonumber\\
&&h=-\frac{(\hat F_{g 2}-F_{g 2}-F_{g1})}{(\hat F_{g 2}-F_{g2}-F_{g1}-3 G_{g 4}
-2 F_{g 3})}\,,
\label{lambda1}
\ea
where $R_1$ is given by,
\be
G_{g3}=G'_{g 3}\,,~~~G_{g4}=G'_{g4}\,.
\ee
Similarly,
\ba
&&\Lambda^*_2=\Big(0,2/(\hat F_{\alpha 2}-F_{\alpha 2}-3 G_{\alpha 4}
+2 G_{2}-2F_{\alpha g3}); R_2\Big)\,,
\nonumber\\
&&h=-\frac{(\hat F_{\alpha 2}-F_{\alpha 2})}{(\hat F_{\alpha 2}
-F_{\alpha 2}-3 G_{\alpha 4}+2 G_{2}
-2 F_{\alpha g 3})}\,,
\label{lambda2}
\ea
where $R_2$ is given by,
\be
G_{\alpha3}=G'_{\alpha 3}\,,~~~G_{\alpha 4}=G'_{\alpha 4}\,.
\ee
And lastly, $\Lambda^*_3(g^{*2}\ne 0, \alpha^* \ne 0)$ is given by,
\ba
g^{*2}&=&2(\hat F_{\alpha2}+G'_{\alpha4}-3G_{\alpha3}+G'_{\alpha3})
\nonumber \\
&&/\big\{(\hat F_{\alpha2}+G'_{\alpha4}-3G_{\alpha3}+G'_{\alpha3})
(4\hat F_{g2}-F_{g1}-F_{g2}-6G_{g3}-2F_{g3})
\nonumber \\
&&-(\hat F_{g2}+G'_{g4}-3G_{g3}+G'_{g3})
(4\hat F_{\alpha2}-F_{\alpha2}+3G_2-6G_{\alpha3}-2F_{\alpha g3})\big\}\,,
\nonumber\\
\alpha^*&=&2(\hat F_{g2}+G'_{g4}-3G_{g3}+G'_{g3})
\nonumber \\
&&/\big\{(\hat F_{g2}+G'_{g4}-3G_{g3}+G'_{g3})
(4\hat F_{\alpha 2}-F_{\alpha 2}+3 G_{2}-6 G_{\alpha 3}-2F_{\alpha g3})
\nonumber \\
&&-(\hat F_{\alpha2}+G'_{\alpha4}-3G_{\alpha3}+G'_{\alpha3})
(4\hat F_{g2}-F_{g1}-F_{g2}-6G_{g3}-2F_{g3})\big\}\,,
\nonumber\\
h&=&-\frac{1}{2}\{(\hat F_{\alpha 2}
-F_{\alpha 2})\alpha^*+(\hat F_{g2}-F_{g2}-F_{g1})g^{*2}\}\,.
\label{lambda3}
\ea

\begin{figure}[htb]
\centering
\includegraphics[width=300pt]{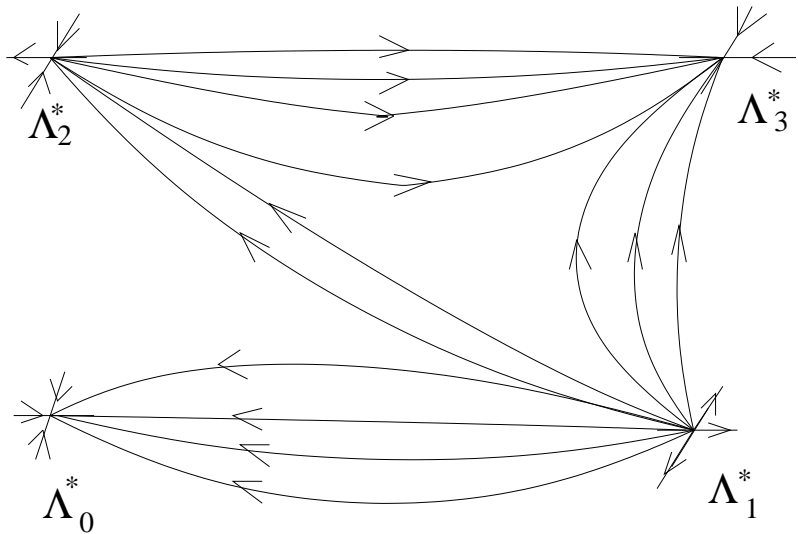} 
\caption{Schematic flow diagram for $R\sim 1_-$ 
in ($g^2$, $\alpha$) plane with the fixed points 
$\Lambda_0^*(0,0)$, $\Lambda_2^*(0,\alpha^*)$, 
$\Lambda_3^*(g^{*2},0)$, $\Lambda_3^*(g^{*2},\alpha^*)$} 
\label{flow}
\end{figure}

\begin{figure}[htb]
\centering
\includegraphics[width=300pt]{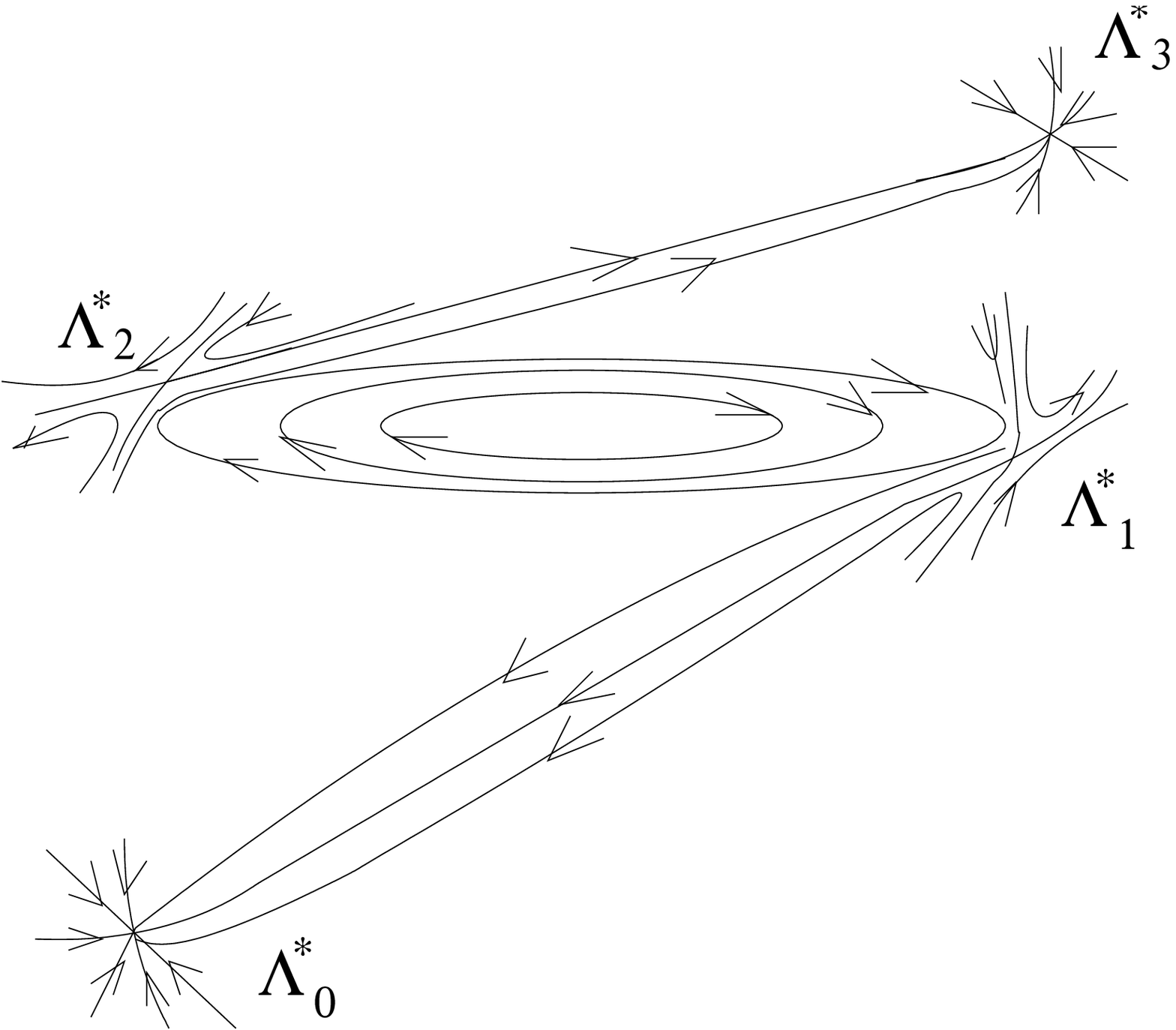} 
\caption{Schematic flow diagram for $R\sim 1_+$ 
in ($g^2$, $\alpha$) plane with the fixed points 
$\Lambda_0^*(0,0)$, $\Lambda_2^*(0,\alpha^*)$, 
$\Lambda_3^*(g^{*2},0)$, $\Lambda_3^*(g^{*2},\alpha^*)$.}
\label{flow1}
\end{figure}

To have a better feeling about the structure of the flow
we plot the flow diagrams for $R \sim 1_{-}$ and  
$R \sim 1_{+}$ in the ($g^2$, $\alpha$) plane (figure \ref{flow}, 
\ref{flow1}), based on the nature of the scaling dimensions we discuss in the next 
sections. As we can extract all the quantities associated with the flow for any radius, 
in principle we can explore the behavior of the flow for a wide range of temperature. 
Remarkably we observe cyclic flow for $R > 1.03$, which is suggestive of resonances of
high spin particles with a stringy behavior \cite{LeClair}. The cyclic flow structure becomes 
very complicated, where the nontrivial fixed points show black hole like behavior. 
Note that a cyclic flow should give periodic $c$-function. However as the multi-critical 
points of matrix models are not necessarily unitary, this does not contradict with
$c$-theorem. 

\subsection{The critical exponents}

Let us now go back to the matrix partition function we started
with. After completing the RG transformations, it obeys the
relation

\be Z_{N+1}[g,\alpha,R-\delta R] \simeq
[\lambda(N,g,\alpha,R-\delta R]^{N^2}~Z_N[g'=g+\delta g,
\alpha'=\alpha+\delta\alpha,R] \,. \ee This leads to the
Callan-Symanzik equation

\be \Big[N \frac{\partial}{\partial N}-\beta_g
\frac{\partial}{\partial g}-\beta_\alpha \frac{\partial}{\partial
\alpha}-\beta_R \frac{\partial}{\partial R}+2 \Big] \mc{F} [
g,\alpha,R ] \approx  r [g, \alpha, R]  \label{CZ} \ee for the
string partition function ({\it i.e.} the world-sheet free energy)

\be \mc{F}[g,\alpha,R] = \frac{1}{N^2}\ln Z[g,\alpha,R] \,. \ee
The singular part of the world-sheet free energy $\mc{F}_s$ is
given by the solution of the homogeneous Callan-Symanzik equation.
The inhomogenous part defined by the change in the prefactor
$\lambda$,
 contributes to subtleties in the free energy.

To discuss the critical exponents for the scaling
variables, the renormalized bulk cosmological constant
$\Delta=1-g/g^*$, and the renormalized fugacity for the vortices
$\hat \alpha=(1-\alpha/\alpha^*)$, we
introduce the scaling dimension matrix

\be \Omega_{k,l} = \frac {\partial \beta_{k}(\Lambda^*)} {\partial
\Lambda_{l} }\,. \ee The eigenvalues $\Omega_i$ of this matrix represent
the scaling dimensions of the relevant operators. The scaling dimensions
at different fixed points are evaluated in the Appendix C.
In terms of them, the homogeneous part of the Callan-Symanzik
equation satisfied by the singular part of the free energy, around a fixed point,
can be written as
\be \Big[N \frac{\partial}{\partial N}-\Omega_{1} ~\Delta
\frac{\partial}{\partial \Delta} -\Omega_{2}~\hat \alpha
\frac{\partial}{\partial~\hat \alpha}+ h~ R \frac{\partial}{\partial R}
+2 \Big] \mc{F}_{s}~[~\Delta,\hat \alpha,R] = 0 \label{cs-scaling}\,.
\ee The scaling of the free energy
with respect to the renormalized cosmological
constant, as one approaches the fixed point, goes as
\be \mc{F} _{s} \sim \Delta ^{2/\Omega_{1}}
F_1 \Big[ N~\Delta^{1/\Omega_1}~, \Delta^{-\Omega_1/\Omega_{2}}
~\hat \alpha~,-\ln \Delta-\Omega_1 \int \frac{dR}{h(R) R}\Big]\,, \ee
where $F_1$ is an arbitrary scaling function whose explicit form
depends on the initial conditions.
Comparing the above expression of $\mc{F}_{s}$
with the matrix model result ~$\mc{F}_{s} \sim \Delta
^{(2-\gamma_{0})} ~f[ N^{2/\gamma_{1}} \Delta]$, or using the
standard definition of the susceptibility ~$\Gamma ~\sim \frac
{\partial^2 \mc{F}_{s}} {\partial \Delta^2} \arrowvert_{\hat \alpha=0}
~\sim \Delta ^{-\gamma_{0}}~$, the string susceptibility exponent
$\gamma_{0}$ is given by

\be \gamma_{0} \sim (2-2/\Omega_{1})\,. \ee Note that in our
analysis $2/\gamma_1 \sim \Omega_{1}$, {\it i.e.} $\gamma_1 \sim
2/\Omega_{1}$ is consistent with the matrix model relation
$\gamma_0 + \gamma_1 = 2$. This relation is independent of the
explicit values of $\gamma_0$ and $\gamma_1$ and is easily
obtainable from the consideration of the torus. The string
susceptibility exponent at genus $G$ is defined by

\be \gamma_G = \gamma_0 + G~\gamma_1 \,.\ee

\subsection{Analysis of the flow}

\begin{figure}[htb]
\centering
\includegraphics[width=250pt,angle=-90]{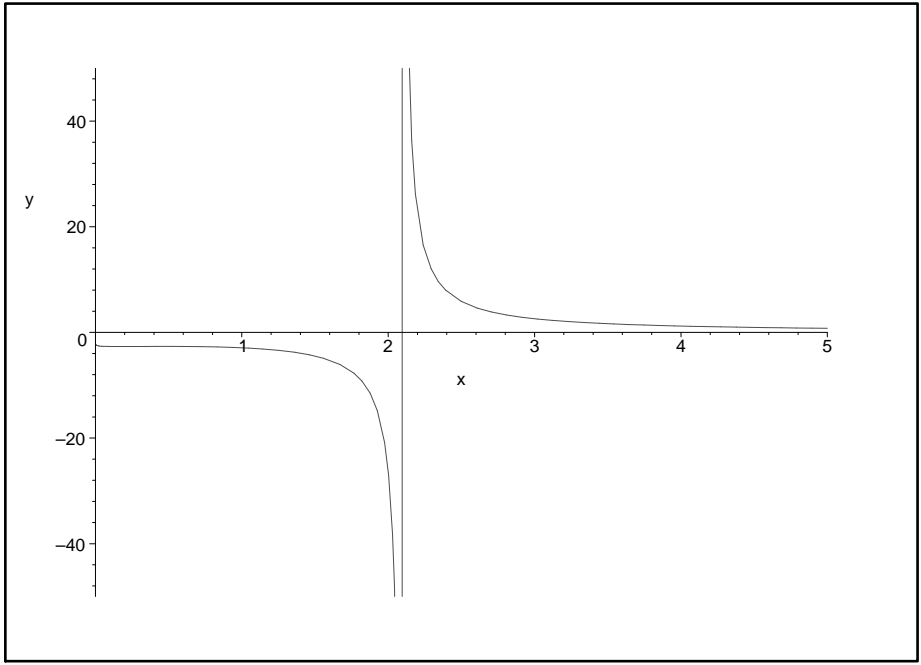} 
\caption{
Behavior of the scaling dimension 
$\Omega_{11}$ for $\Lambda_1^*$ with respect to
$x=\pi R$} \label{O11}
\end{figure}

\begin{figure}[htb]
\centering
\includegraphics[width=250pt,angle=-90]{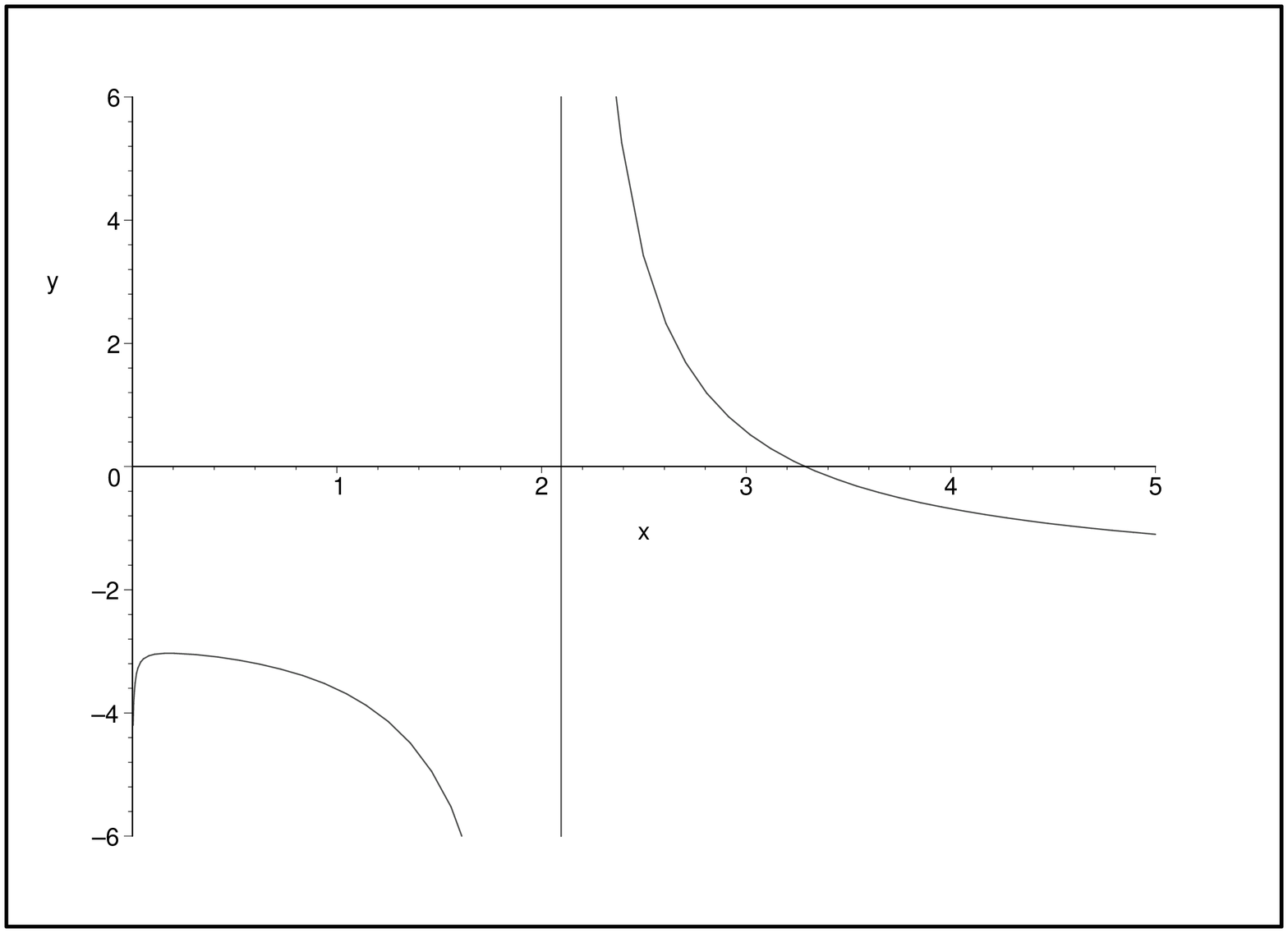}
 \caption{
Behavior of the scaling dimension 
$\Omega_{22}$ for $\Lambda_1^*$ with respect to
$x=\pi R$} \label{O22}
\end{figure}

Now we will analyze the nature of the fixed points considering the
behavior of the scaling dimensions  with respect to the change of
the compactification radius $R$. Since the flow equations and the
constraints together give rise to complicated system of nonlinear
simultaneous equations, it would be easier to analyze them by studying the
behavior of all quantities as functions of temperature. We will also plot the
relevant functions for the convenience of the analysis.
Referring to the Appendix C we
observe that, the nontrivial fixed point $\Lambda_1^*$ produces a
{\it pair of $c=1$ fixed points} for $R \to \infty$, characterized
by the asymptotic scaling dimensions \{$\Omega_1=1,
~\Omega_2=-1.2$\} (more precisely, the asymptotic values are true
for $R \ge 1.08$ and $R \ge 2$ respectively, see figure \ref{O11}, 
\ref{O22}) and hence the string susceptibility exponent
$\gamma_0=(2-2/\Omega_1)=0$. 

These fixed points have one unstable
direction. The critical coupling for such fixed points is
vanishingly small ($g^{*2} \to 0, \alpha^*=0$) and hence the pair
of $c=1$ fixed points will be infinitesimally close to the
gaussian fixed point, as we have already seen  in the analysis of
the ungauged model \cite{dd-singlet}.
As the radius is increased $\Omega_1$ grows to
infinity as $R \to 0.70$, and $\Omega_2$ flips sign from negative
to positive at $R=1.03$ and also grows to infinity as $R \to
0.76$. This indicates that the operator coupled to the fugacity
of vortices becomes relevant at $R=1.03$ and triggers the expected
{\it liberation of the world-sheet vortices  by
BKT transition} at the self dual radius \cite{GK1,GK2}.

\begin{figure}[htb]
\centering
\includegraphics[width=250pt,angle=-90]{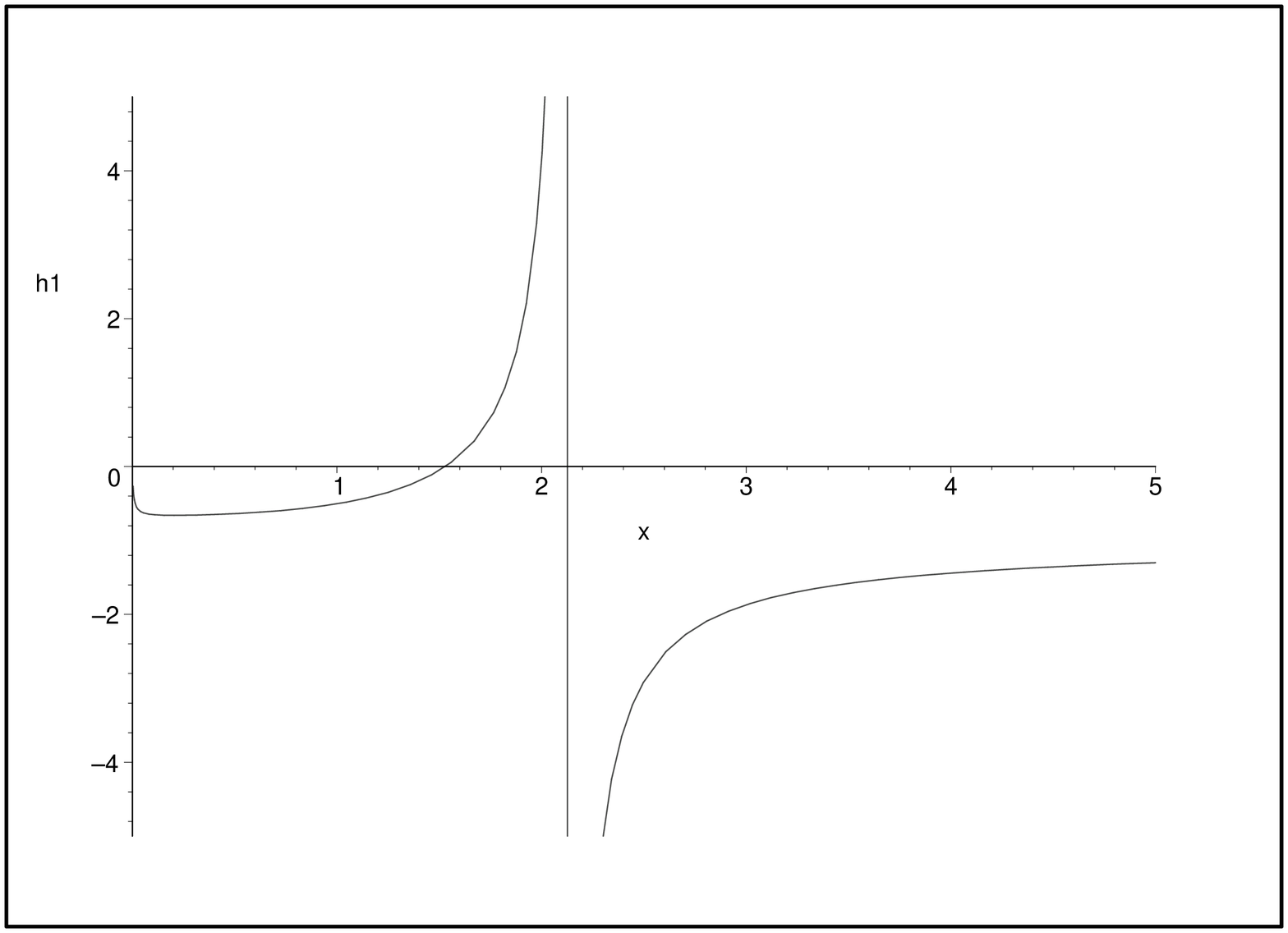} 
\caption{
Behavior of $h$ for $\Lambda_1^*$ with respect to
$x=\pi R$} \label{h1}
\end{figure}

After this transition, the fixed point becomes purely repulsive up
to $R \sim 0.67$.  In the range $0.67\le R \le 0.70$, the critical
coupling  is infinitely large $(g^{*2} \to \infty , \alpha^* =0)$
as well as the scaling exponents $\Omega_1 \to \infty, \Omega_2
\to \infty$. As the quantity $h$ is also a large positive quantity
here (figure \ref{h1}), the fixed point $\Lambda_1^*$ in this region is
characterized by a negative specific heat, reminiscent of
{\it Euclidean black hole} in flat space-time. We will elaborate on
this in the next section. These black hole like fixed points have a
positive string susceptibility exponent $\gamma_0 =(2-2/\Omega_1)
\to 2$.

Between $0 < R \le 0.67$  the matrix coupling $g^{*2}$, $\alpha^*$
and both of the scaling dimensions are  negative. Hence one gets a
purely attractive fixed point with imaginary matrix coupling.
However, as $R \to 0$, the couplings are vanishingly small, also
$h \to 0$, and one reaches {\it a pure gravity fixed point} ($c=0$)
with one unstable direction as $\Omega_1 \sim 0.8$ (such that
$\gamma_0=2-2.5 \sim -0.5$) and $\Omega_2 \to -3.6$.
Note that even though $\Lambda_1^*$ exhibits so many
important features for a wide range of temperature,
the constraint we have here
allows us to look at it only around $R=1$ and around $R=0$.
We need to improve over this constraint.

\begin{figure}[htb]
\centering
\includegraphics[width=250pt,angle=-90]{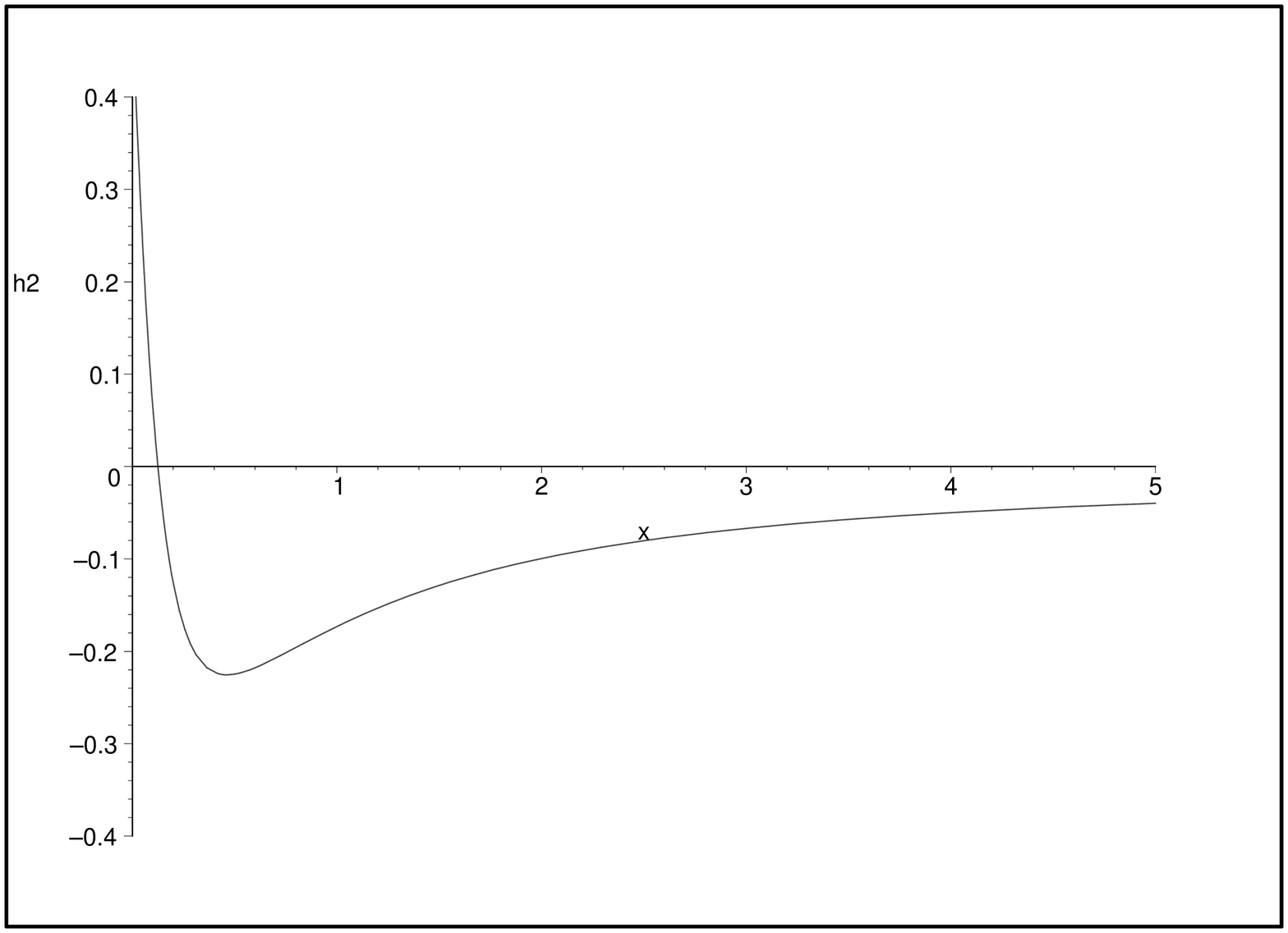} 
\caption{
Behavior of  
$h$ for $\Lambda_2^*$ with respect to
$x=\pi R$} \label{h2}
\end{figure}

\begin{figure}[htb]
\centering
\includegraphics[width=250pt,angle=-90]{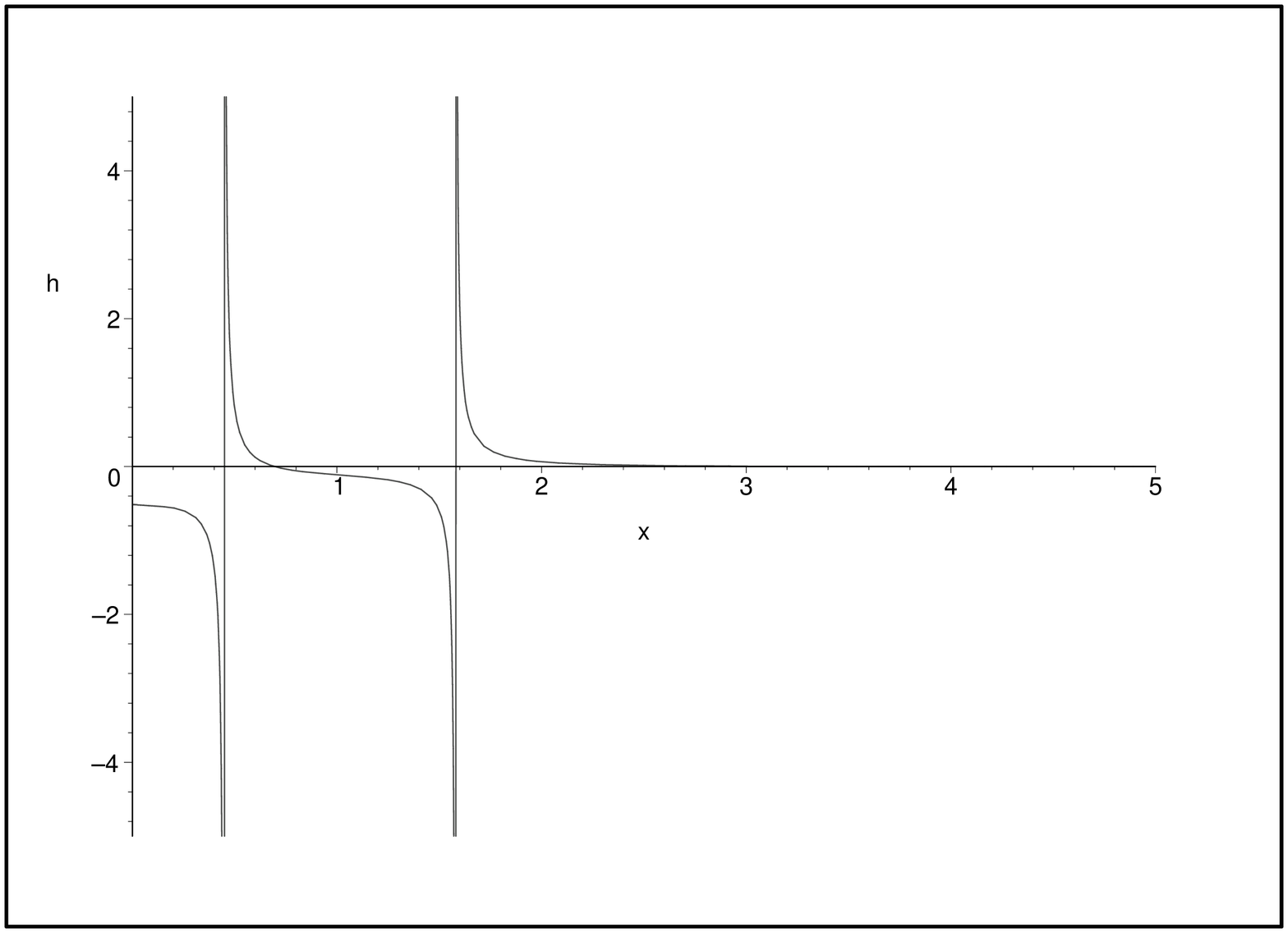} 
\caption{
Behavior of 
$h$ for $\Lambda_3^*$ with respect to
$x=\pi R$} \label{h3}
\end{figure}

Similarly one can analyze the behavior of the fixed point
$\Lambda_2^*$ with respect to the parameter $R$. It exhibits {\it
black hole like behavior} ($h$ is
large positive quantity) as
$ 0 \le R \ge 0.25
0$. In this limit the fixed point $\Lambda_2^*(g^*=0, \alpha^*
\to \infty)$ is purely repulsive, $\Omega_1\to \infty, \Omega_2
\to \infty$, and hence $\gamma_0 \to 2$ 
(see figure \ref{h2}). However, here
the constraint on the radius is $R \ge 0.19$.
For the fixed points $\Lambda_3^*$, there is also similar
black hole like behavior with negative specific heat
at $0.51\le R \ge 0.63$ and at $0.12\le R\ge 0.19$ and
$\gamma_0=2$ (see figure \ref{h3}). 
There is no constraint on the radius.

\section{Comments On The Black Hole Fixed Point}
\setcounter{equation}{0}

To study the thermodynamic behavior of the fixed points we would
like to solve for the free energy from the Callan-Symanzik
equation as one approaches the BKT phase transition (or the
Hagedorn transition) and analyze its thermal properties. From the
analysis of the previous section, one can see that the fixed point
$\Lambda_1^*$ has rather exotic behavior around $R_H \sim 0.73$,
little below (above) the Kosterlitz-Thouless radius (temperature)
$R_{KT}=1/(2 \pi T_{KT})=1$. Around this region the scaling
dimensions $\Omega_1$, $\Omega_2$, and $h$, are large constants
which simplifies the situation. However, as we proceed we will see
that the black hole like behavior would emerge from the region
where $h$ is a large positive constant. Thus around $R_H$ the free
energy can be written as

\be
\mc{F}_s \sim f[(R-R_H/R_H)^{1/h}, \Delta, \hat \alpha, N]\,,
\ee
where the inverse temperatures are defined by,

\be
\beta = 2\pi R = \frac{1}{T}\,,~~~~\beta_H = 2\pi R_H = \frac{1}{T_H}\,.
\ee
The thermodynamic quantities are given by

\ba
\mc{F}_s (\beta-\beta_H) &=& -\beta~F(\beta-\beta_H) =
\ln Z(\beta-\beta_H) \,,
\nonumber \\
E &=& \frac{\partial(\beta F)}{\partial\beta} = \frac{-\partial
\ln Z}{\partial\beta} \,,
\nonumber \\
C_v &=&-\beta^2 \Big(\frac{\partial E}{\partial \beta}\Big)_v\,.
\ea
Using $\mc{F}_s$, we have

\ba
E &\sim& \frac{1}{h}~\beta_H^{-1}~\Big(\frac{\beta-\beta_H}
{\beta_H}\Big)^{1/h-1}\,.
\nonumber \\
\label{energy} \ea Since near the phase transition the fluctuation
of energy is large, the canonical ensemble would diverge. In such
situation, it is better to pass to the microcanonical ensemble
with fixed energy and the temperature defined by

\be \beta = \frac{\partial S(E)}{\partial E} \,. \label{microtemp}
\ee Using (\ref{energy}) for large positive $h$, one can solve for
$\beta$ in terms of $E$ as \be \beta-\beta_H \sim
-\frac{h^{-1}}{E}\,. \label{bh-entropy} \ee Combining this with
the definition of temperature in the microcanonical ensemble
(\ref{microtemp}) one can calculate the near Hagedorn one loop
finite energy  correction to the usual definition of Hagedorn
density of states, $\rho (E) = \exp[S(E)] \sim \exp[\beta_H E]$,
and the usual inverse temperature, which is otherwise a constant
$\partial S/\partial E = \beta_H$. The finite energy corrections
are of the form

\ba \beta &=& \frac{\partial\ln\rho}{\partial E} = \beta_H +
\frac{s_1}{E}  + O\bigg(\frac{1}{E^2}\bigg) \,,
\nonumber \\
\rho(E) &\sim& E^{s_1}~\exp[\beta_H
E]\bigg[1+O\bigg(\frac{1}{E}\bigg)\bigg] \,. \label{oneloopcorr}
\ea Here the number $s_1$ would come from the one loop correction. If
$s_1$ is negative, the specific heat is negative, {\it i.e.}
increasing the energy of the system gives rise to the decrease of
temperature, indicating {\it Euclidean black hole like behavior in flat
spacetime}.

\begin{figure}[htb]
\centering
\includegraphics[width=250pt]{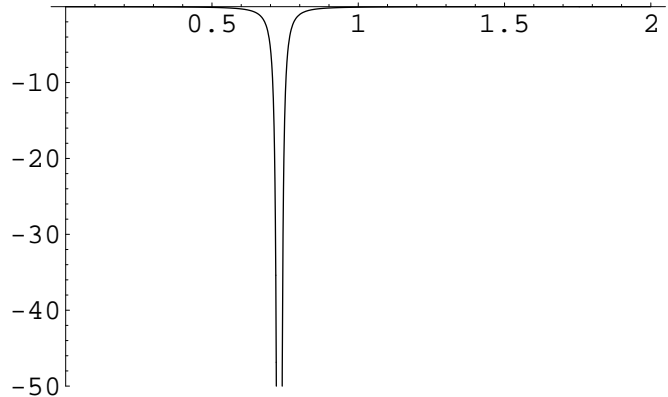} 
\caption{
Behavior of the specific heat $C_v$ (for $\Lambda_1^*$) 
with respect to $x=\pi R$} \label{Cv1}
\end{figure}

Integrating our equation (\ref{bh-entropy}) to get the
entropy and density of states
we identify the one-loop correction as
\ba
S(E) \sim \beta_H E-\frac{1}{h} \ln E\,,
\nonumber\\
\rho(E) \sim E^{-\frac{1}{h}} \exp [\beta_H E]\,,~~~s_1=-1/h <
0\,. \label{entropy-dos} \ea This behavior is true for large
positive $h$ and the corresponding range of radius only. As we
have analyzed in the previous section, we encounter such fixed
points $\Lambda_1^*$ of very large and positive $h$ in the region
$0.67 \le R \le 0.70$. These are pair of purely repulsive fixed
points (over this region the scaling exponents $\Omega_1,
\Omega_2$ are also large positive constants like $h$) of large
(diverging) coupling $g^{*2} \to \infty, \alpha^*=0$ and positive
string susceptibility exponent $\gamma_0 \sim 2$. We therefore
identify such fixed points as the {\it Euclidean Black hole like
fixed points} in the continuum limit, with a negative specific
heat (figure \ref{Cv1})

\be
C_v \sim -\frac{1}{h}~\frac{\beta^2}{(\beta-\beta_H)^{2}} \,.
\label{sp-heat}
\ee
Using relations like $C_v=- \beta \Big(\frac{\partial S}
{\partial \beta}\Big)_v$
the entropy as a function of temperature is

\be S(\beta-\beta_H) \sim -\frac{1}{h}\Big(\log (\beta-\beta_H)-
\frac{\beta_H}{\beta-\beta_H}\Big)\,. \label{entropy-thermal} \ee
The discontinuity in entropy (a measure of the latent heat of the
transition) suggest that the Hagedorn transition is a first order
phase transition at little higher temperature than the KT
temperature, driving the system to an unstable black hole phase.

\section{Discussions}
\setcounter{equation}{0}

In this paper, starting from a matrix quantum mechanics on a
circle with a periodic boundary condition (\ref{AZ1N}) (with gauge
fields $A(t)=\Omega^\dagger \dot\Omega (t)$ as degrees of freedom
apart from the matrix degrees of freedom $\phi(t)$ and also with
the coupling $\alpha$ as the fugacity of vortices), we have
analyzed the phase structure of the theory by a world-sheet
renormalization group flow and observed following remarkable
facts. The nontrivial fixed points of the flow does capture the
known physics of the $2D$ string theory, namely a fixed point with
the critical index of $c=1$ for large radius ($R>1$), and moreover
reveals a new phase. In our previous paper
\cite{dd-singlet} we have analyzed this
$c=1$ fixed point in detail showing that it exhibits expected
logarithmic scaling violation of the singlet free energy and the
T-duality. Note that for this class of fixed point
($\Lambda_1^*$), a pair with $\alpha^*=0$, the vanishing $\alpha$
forces $A$ to be zero and the matrix quantum mechanics
(\ref{AZ1N}) reduces to the usual matrix quantum mechanics with
periodic boundary condition studied in \cite{dd-singlet}\footnote{Working with
(\ref{AZ1N}) and then having $\alpha^*=0$ seems to improve the
critical exponent in studying $c=1$ fixed point with a nonzero $h$.
However, we still need to improve on the constraint for
$\Lambda_1^*$ which allows us to look at this particular fixed
point around $R=1$ and around $R=0$ only.}. The flow equations
also exhibit indication of phase transition at $R=1.03$ as
expected in BKT phase transition undergone by the $2D$ string
theory due to liberation of world-sheet vortices. However, a new
phase emerges beyond the phase transition at the self-dual radius.
The fixed points $\Lambda_1^*$ exhibit an unstable black hole like
phase with negative specific heat, between $0.67<R<0.7$. As $R \to
0$ they end up into $c=0$ fixed points which is consistent with
the expectation. For another class of fixed points $\Lambda_3^*$,
a pair with both nonzero $\alpha^*$ and $g^*$, there is unstable
black hole like behavior with negative specific heat above the
self-dual temperature (at $0.51<R<0.63$ and at
$0.12<R<0.19$)\footnote{Though for $\Lambda_2^*$, a pair of fixed
points with nonzero $\alpha^*$, the behavior is seen for radii
$0<R<0.25$, there is a constraint to get the fixed point which
renders $R \ge 0.19$.}. From a thermodynamic study of the free
energy obtained from the Callan-Symanzik equations we show that
all these unstable phases do have negative specific heat. The
thermodynamic quantities indicate that the system does undergo a
finite temperature phase transition around the Hagedorn
temperature ($R_H=0.7$, around which the new phase is formed) and
exhibits one loop finite energy correction to the Hagedorn density
of states.

Thus the thermodynamics of the $2D$ string theory above the BKT
phase transition at the self-dual radius is governed by the
unstable black hole like phase with negative specific heat for a
range of temperatures beyond the BKT phase transition point. The
remarkable thing is that we observe phases of negative specific
heat for a range of temperatures rather than at a particular
temperature like the case in \cite{KKK} indicating that
presumably there are many black holes at temperatures other than
that of $2D$ {\it cigar black hole} described by
dilaton gravity. It is then meaningful to ask where
do the thermal black holes that we observe end up if they
evaporate? Our analysis shows one class of these
unstable objects evaporate to $c=0$, the
other classes end up in other black holes at higher
temperatures. In $1+1$ dimension, the string theory is integrable
due to infinite number of conserved charges. Thus from continuum
point of view it is possible that there are other black hole
solutions not only characterized by mass and temperature but also
by other values of conserved charges, in which case they might be
at different temperatures. Here we can mention that the
understanding that the integrability of the free fermion structure
perhaps prevents the formation of $c=1$ black hole is consistent
from the nature of the flow and the fixed points. The $c=1$ fixed
points at large $R$, dominated by the singlet sector, does not
exhibit any black hole like behavior.

To actually see that these objects are black holes and to deal
with the questions like the formation and the unitarity of the
scattering off the black hole one needs to study the dynamics of
the nonsinglet states. As the observations indicate the
nonsinglet states account for the entropy of the black hole, it
would be nice to realize them as the excitations of the black
hole. In black hole physics it was proposed that presumably the
black hole entropy is due to open string with ends lying on the
event horizon. It would be really interesting to study the
nonsinglet boundaries of the matrix quantum mechanics in this
context. The general framework of the world-sheet renormalization
group approach is useful to study these objects, which is
otherwise difficult. The Hamiltonian formalism, appropriate to
address the questions of black hole dynamics in the Minkowski
space, is complicated due to Calogero type of interaction.
However, in the renormalization group approach one can utilize the
Callan-Symanzic equations to calculate the wave functions for the
nonsinglet boundaries or the macroscopic loops and can construct
a $S$-matrix. The wave equation for the macroscopic loops should
then contain the information regarding the metric of the black
hole to "see" the black hole at all. In this regard one
nice observation from the renormalization group analysis
is that the running of the radius with scale
implies a deformation in the target space geometry if one considers
the scale to go like dilaton. Presumably this could help to illuminate
further the issue of getting the metric. Also it is interesting to
understand the localized wave function (microscopic loop)
describing the the tip of the cigar black hole \cite{SeibergShenker}.
The unique boundary associated to the inner core of the black hole
is thought to be essential to understand the evaporation and the
Hawking radiation \cite{HoroMalda}. We have work in progress
on understanding the boundaries in this context.

Regarding our RG method, as we have already discussed in
\cite{dd-singlet}, it would be interesting to generalize the scheme for arbitrary
couplings and to keep arbitrary powers of $\phi$ in the evaluation
of the determinant obtained from the integration over the vector
degrees of freedom ($v, v^*$). Here we will just mention that (as discussed
in our previous papaer) because of these vectors our partition function (in
the $\alpha\to 0$ limit) essentially looks like the model discussed
in \cite{Yang,Minahan} and thus is useful
to understand the presence of the boundaries. As a simple step towards 
generalization we would like to study the flow with an additional 
coupling due to $\phi^4$ term. These might reveal finer 
observations and also would be helpful to test the convergence of the
scheme as well. The cyclic flow structure also deserves a detail study, 
specially near the regime of the black hole like behavior of the nontrivial 
fixed points where the flow structure is complicated.  
We would also like to understand how the relevant operators
driving our flow look like in the matrix quantum mechanics side
and how they translate to the operators in the world-sheet.

\vskip 1.0cm

{\bf Acknowledgments}

We would like to thank Michael Douglas for useful discussions,
comments and constant support at all stages of the work. We would
also like to thank Igor Klebanov, Massimo Porrati for discussions
and especially Edouard Br\'ezin for illuminating discussions,
advices and reading an early draft. We thank the organizers of the
string workshop at HRI, India for hospitality during early stages
of this work where part of this work was reported. The research of
SD was supported in part by DOE grant DE-FG02-96ER40959. The
research of TD was supported in part by NSF grants PHY-0070787 and
PHY-0245068. Any opinions, findings, and conclusions or
recommendations expressed in this material are those of the
authors and do not necessarily reflect the views of the National
Science Foundation.


\addcontentsline{toc}{section}{Appendix}
\appendix{The expression of $\Sigma [g,\alpha,\phi_N,A_N,R,N]$}\label{A1}

\newpage
Evaluating the diagrams, $\Sigma [g,\alpha,\phi_N,A_N,R,N]$ can be
expressed as

\ba &&\Sigma[g,\alpha,\phi_N,A_N,R,N] = 1 +
i~\mbox{Tr}\bigg[\sum_n \frac{2n/R} {\frac{n^2}{R^2}+1}~A_0\bigg]
+ g~\mbox{Tr}\bigg[\sum_n\frac{1} {\frac{n^2}{R^2}+1}~\phi_0\bigg]
+ \alpha~\mbox{Tr}\bigg[-\sum_k\phi_k\phi_{-k}
\nonumber \\
&&+\sum_{m,n}\frac{(2m^2-5mn+2n^2)/R^2}{\frac{m^2}{R^2}+1}~\phi_{m-n}\phi_{n-m}
\bigg] +
\frac{g^2}{2}~\mbox{Tr}\bigg[\sum_{m,n}\frac{1}{\big(\frac{m^2}{R^2}+1
\big)\big(\frac{n^2}{R^2}+1\big)}~\phi_{m-n}\phi_{n-m}\bigg]
\nonumber \\
&&-\mbox{Tr}\Bigg[\sum_{n,k}\frac{1}{\frac{n^2}{R^2}+1}~A_kA_{-k}
+\frac{1}{2!}\sum_{m,n}\frac{(m^2+2mn+n^2)/R^2}{\big(\frac{m^2}{R^2}+1
\big)\big(\frac{n^2}{R^2}+1\big)}~A_{m-n}A_{n-m}\Bigg]
\nonumber \\
&&+i\alpha~\mbox{Tr}\Bigg[\sum_{m,n,k}
\bigg\{\frac{(4n-2m)/R}{\frac{m^2}{R^2}+1}
-\frac{(2n-m)/R}{\frac{n^2}{R^2}+1}\bigg\}~A_{m-n-k}\phi_k\phi_{n-m}
\nonumber \\
&&+\sum_{m,n,k} \bigg\{\frac{(4n-2m)/R}{\frac{n^2}{R^2}+1}
-\frac{(2n-m)/R}{\frac{m^2}{R^2}+1}\bigg\}~A_{m-n-k}\phi_{n-m}\phi_k
\nonumber \\
&&-\sum_{m,n,l}\frac{(2m^2l+4ml^2+m^2n-2l^2n-2m^3-mnl)/R^3}
{\big(\frac{m^2}{R^2}+1\big)\big(\frac{l^2}{R^2}+1\big)}
~A_{m-l}\phi_{l-n}\phi_{n-m}\Bigg]
\nonumber \\
&&+i\frac{g^2}{2}~\mbox{Tr}\Bigg[\sum_{m,n,l}\frac{(m+l)/R}
{\big(\frac{m^2}{R^2}+1\big)\big(\frac{n^2}{R^2}+1\big)
\big(\frac{l^2}{R^2}+1\big)}~A_{m-l}\phi_{l-n}\phi_{n-m}\Bigg]
\nonumber \\
&&-\frac{g^3}{3!}~\mbox{Tr}\Bigg[\sum_{m,n,l}\frac{1}
{\big(\frac{m^2}{R^2}+1\big)\big(\frac{n^2}{R^2}+1\big)
\big(\frac{l^2}{R^2}+1\big)}~\phi_{m-l}\phi_{l-n}\phi_{n-m} \Bigg]
\nonumber \\
&&-\alpha g~\mbox{Tr}\Bigg[\sum_{m,n,l}\frac{(4ml-2mn-2ln+n^2)/R^2}
{\big(\frac{m^2}{R^2}+1\big)\big(\frac{l^2}{R^2}+1\big)}
~\phi_{m-l}\phi_{l-n}\phi_{n-m}\Bigg]
\nonumber \\
&&+\alpha~\mbox{Tr}\Bigg[ \sum_{m,n,k,k'}
\bigg\{\frac{4}{\frac{m^2}{R^2}+1}+\frac{1}{\frac{n^2}{R^2}+1}\bigg\}
~A_{m-n-k}\phi_k\phi_{k'}A_{n-m-k'}
\nonumber \\
&&-\sum_{m,n,k,k'}\frac{2}{\frac{m^2}{R^2}+1}
(\phi_kA_{m-n-k}\phi_{k'}A_{n-m-k'}+\phi_kA_{n-m-k'}\phi_{k'}A_{m-n-k})
\nonumber \\
&&-\sum_{m,n,l,k}\frac{(2m^2-nm+2ml-nl)/R^2}{\big(\frac{m^2}{R^2}+1\big)
\big(\frac{l^2}{R^2}+1\big)}~(2A_{l-n-k}\phi_k\phi_{n-m}A_{m-l}-\phi_kA_{l-n-k}
\phi_{n-m}A_{m-l})
\nonumber \\
&&+\sum_{m,n,l,k}\frac{(2l^2-nm+2ml-nl)/R^2}{\big(\frac{m^2}{R^2}+1\big)
\big(\frac{l^2}{R^2}+1\big)}~(2\phi_{l-n}\phi_kA_{n-m-k}A_{m-l}
-\phi_{l-n}A_{n-m-k}\phi_kA_{m-l})
\nonumber \\
&&+\sum_{m,n,l,k}\frac{(n^2-2nm+4ml-2nl)/R^2}{\big(\frac{m^2}{R^2}+1\big)
\big(\frac{l^2}{R^2}+1\big)}~2\phi_{l-n}\phi_{n-m}A_{n-l-k}A_k\Bigg]
\nonumber \\
&&-\frac{g^2}{2}~\mbox{Tr}\Bigg[\sum_{m,n,l,k}
\frac{1}{\big(\frac{m^2}{R^2}+1\big)\big(\frac{n^2}{R^2}+1\big)
\big(\frac{l^2}{R^2}+1\big)}~\phi_{l-n}\phi_{n-m}A_{n-l-k}A_k
\Bigg]\,. \label{SigmaFourier} \ea


\appendix{The Feynman Diagrams}\label{A2}

Here we evaluate and discuss the terms in different orders of the
series $\Sigma [g,\alpha,\phi_N,A_N,R,N]$ (\ref{SigmaFourier})
using the summation rules discussed in section (3.3) and the
relation (\ref{FourierTrfm}) for the inverse Fourier Transform.

\subsection {The terms of order O($\Phi$):}

\be g~\mbox{Tr}\bigg[\sum_n\frac{1}
{(\frac{n^2}{R^2}+1)}~\phi_0\bigg]=g \coth (\pi R) \int_{0}^{2 \pi
R}dt ~\mbox{Tr}\Phi(t)\,.
 \ee

\subsection {The terms of order O($A$):}

\be
i~\mbox{Tr}\bigg[\sum_n \frac{2n/R} {(\frac{n^2}{R^2}+1)}~A_0\bigg]=0\,.
\ee

This is because,

$$
\sum_n \frac{n/R}{(\frac{n^2}{R^2}+1)}=0\,.
$$

\subsection {The terms of order O($\Phi$$\Phi$):}

\ba
(1)&&\frac{g^2}{2}~\mbox{Tr}\bigg[\sum_{m,n}\frac{1}{\big(\frac{m^2}{R^2}+1
\big)\big(\frac{n^2}{R^2}+1\big)}~\phi_{m-n}\phi_{n-m}\bigg]
\nonumber\\
&&=\frac{g^2}{2} \int \frac{dt_{1} dt_{2}}{(2 \pi R)^2}
~\mbox{Tr}( \Phi(t_{1})\Phi(t_{2}))
\bigg[\sum_{m,n}\frac{\exp\big(i(n-m)~t_{1}/R\big)
\exp\big(i(m-n)~t_{2}/R\big) }{\big(\frac{m^2}{R^2}+1
\big)\big(\frac{n^2}{R^2}+1\big)} \bigg]\,.
\nonumber\\
\ea Now, changing the variables to 'center of mass' and 'relative'
coordinates defined respectively by \be
T=\frac{t_{1}+t_{2}}{2}\,,~~~~~\tau=\frac{t_{1}-t_{2}}{2}\,, \ee
we have, \be dt_{1}~dt_{2}=    J  \Big(\frac{t_{1},t_{2}}{T,
\tau}\Big)~dT~d\tau=2~dT~d\tau\,. \ee 
Hence, \ba
&&\frac{g^2}{2}~\mbox{Tr}\bigg[\sum_{m,n}\frac{1}{\big(\frac{m^2}{R^2}+1
\big)\big(\frac{n^2}{R^2}+1\big)}~\phi_{m-n}\phi_{n-m}\bigg]
\nonumber\\
&&=\frac{g^2}{4 \pi^2 R^2} \int_{0}^{2 \pi R} dT \int_{-\pi
R}^{\pi R}d\tau ~\mbox{Tr}( \Phi\big(T+\tau)\Phi(T-\tau)\big)
\sum_{m,n}\frac{\exp\big(i(n-m)~2\tau/R\big)}{\big(\frac{m^2}{R^2}+1
\big) \big(\frac{n^2}{R^2}+1\big)}
\nonumber\\
&&\simeq \frac{g^2}{4 \pi^2 R^2} \int_{0}^{2 \pi R} dT \int_{-\pi
R}^{\pi R}d\tau~\mbox{Tr}( \Phi(T)^2-\tau^2 \dot \Phi(T)^2)
\sum_{m,n}\frac{\exp\big(i(n-m)~2\tau/R\big)}{\big(\frac{m^2}{R^2}+1
\big)\big(\frac{n^2}{R^2}+1\big)}
\nonumber\\
&& \simeq g^2 F_{g2} (R) \int_{0}^{2 \pi R} dT~\mbox{Tr}
(\Phi(T)^2/2)+g^2\hat F_{g2} (R) \int_{0}^{2 \pi R} dT
~\mbox{Tr}(\dot \Phi(T)^2/2)\,, \ea where, \ba F_{g2}
(R)&=&\frac{1}{2 \pi^2 R^2}\int_{-\pi R}^{\pi R} d\tau \sum_{m,n}
\frac{ \exp\big(i(m-n)~2\tau/R\big)}{\big(\frac{m^2}{R^2}+1 \big)
\big(\frac{n^2}{R^2}+1\big)}
\nonumber\\
&=&\frac{1}{2 \pi^2 R^2}\int_{0}^{2 \pi R} d\tau' \sum_{m,n}
\frac{\exp\big(i(m-n)~\tau'/R\big)}{\big(\frac{m^2}{R^2}+1 \big)
\big(\frac{n^2}{R^2}+1\big)}\,,~~~~(\tau'=2 \tau)
\nonumber\\
&=&\frac{1}{2 \sinh^2 \pi R} \int _{0}^{2 \pi R} d\tau'\cosh (\pi
R+  \tau') \cosh(\pi R- \tau')
\nonumber\\
&=&\frac{1}{\sinh^2 \pi R} \bigg[\frac{1}{2} \pi R\cosh 2\pi
R+\frac{1}{8}\sinh 4\pi R\bigg]\,, \ea and, \ba \hat
F_{g2}(R)&=&-\frac{1}{2 \pi^2 R^2}\int_{-\pi R}^{\pi R}
d\tau~\tau^2 \sum_{m,n} \frac{
\exp\big(i(m-n)~2\tau/R\big)}{\big(\frac{m^2}{R^2}+1 \big)
\big(\frac{n^2}{R^2}+1\big)}
\nonumber\\
&=&-\frac{1}{8 \pi^2 R^2}\int_{o}^{2 \pi R}
d\tau'~\tau'^2\sum_{m,n}\frac{
\exp\big(i(m-n)~\tau'/R\big)}{\big(\frac{m^2}{R^2}+1 \big)
\big(\frac{n^2}{R^2}+1\big)} \,,~~~~(\tau'=2 \tau)
\nonumber\\
&=&-\frac{1}{8 \sinh^2 \pi R}\int_{o}^{2 \pi R}
d\tau'~\tau'^2\cosh(\pi R+\tau) \cosh(\pi R-\tau)
\nonumber\\
&=& \frac{1}{\sinh^2 \pi R}\bigg[\frac{1}{64}(1+8 \pi^2 R^2)\sinh
4\pi R -\frac{\pi^3 R^3}{6}\cosh 2\pi R-\frac{\pi R}{16} \cosh 4
\pi R \bigg]\,.
\nonumber\\
\ea 
\ba 
(2)&&\alpha~\mbox{Tr}\bigg[\sum_{m,n}
\frac{(2m^2-5mn+2n^2)/R^2}{(\frac{m^2}{R^2}+1)}~\phi_{m-n}\phi_{n-m}
\bigg]=\alpha \int \frac{dt_{1} dt_{2}}{(2 \pi R)^2}
~\mbox{Tr}(\Phi(t_{1})\Phi(t_{2}))
\nonumber\\
&&\bigg[\sum_{m,n}\frac{(2m^2-5mn+2n^2)/R^2}{(\frac{m^2}{R^2}+1)}
\exp\big(i(n-m)~t_{1}/R\big)\exp\big(i(m-n)~t_{2}/R\big)
\bigg]
\nonumber\\
&&=\frac{\alpha}{2 \pi^2 R^2} \int_{0}^{2 \pi R} dT\int_{-\pi
R}^{\pi R} d\tau  ~\mbox{Tr}\big( \Phi(T+\tau) \Phi(T-\tau)\big)
\nonumber \\
&&\bigg[\sum_{m,n}\frac{(2m^2-5mn+2n^2)/R^2}{(\frac{m^2}{R^2}+1)}\exp\big(i(n-m)~2
\tau/R\big) \bigg]
\nonumber\\
&& \simeq \frac{\alpha}{2 \pi^2 R^2} \int_{0}^{2 \pi R}
dT\int_{-\pi R}^{\pi R} d\tau  ~\mbox{Tr}( \Phi(T)^2-\tau^2 \dot
\Phi(T)^2)
\nonumber \\
&&\bigg[\sum_{m,n}\frac{(2m^2-5mn+2n^2)/R^2}{(\frac{m^2}{R^2}+1)}\exp\big(i(n-m)~2
\tau/R\big) \bigg]
\nonumber\\
\ea 
Now, \ba &&\frac{1}{2 \pi^2 R^2}\int_{-\pi R}^{\pi R}
d\tau\bigg[\sum_{m,n}\frac{(2m^2-5mn+2n^2)/R^2}{(\frac{m^2}{R^2}+1)}\exp\big(i(n-m)~2
\tau/R\big)\bigg]
\nonumber\\
&&= \frac{1}{4 \pi^2 R^2}\int_{0}^{2 \pi R}
d\tau'\sum_{m,n}\frac{(2m^2-5mn+2n^2)/R^2}{(\frac{m^2}{R^2}+1)}
\nonumber \\
&&\bigg[\exp\big(i(n-m)~\tau'/R\big)+\exp\big(-i(n-m)~\tau'/R\big)\bigg]
\nonumber\\
&&= \int_{0}^{2 \pi R} d\tau' \bigg[4
\delta(-\tau')~\delta(\tau')-5 \cosh \tau~\delta'(\tau')-2
\coth \pi R~ \cosh \tau~\delta''(\tau') \bigg]\,,~~~(\tau'=2 \tau)
\nonumber\\
&&= \bigg[4-5~\lim_{\epsilon \rightarrow 0} \frac{\cosh
(\epsilon)-\cosh (-\epsilon)}{2 \epsilon}
-2 \coth \pi R~\lim_{\epsilon_{1} \rightarrow 0}
\frac{\sinh \epsilon_{1}}{\epsilon_{1}} ~\lim_{\epsilon_{2}
\rightarrow 0} \frac{\sinh \epsilon_{2}}{\epsilon_{2}} \bigg] 
\nonumber \\
&&=(4-2 \coth \pi R)\,,
\nonumber\\
\ea
and,
\ba
&&\frac{1}{2 \pi^2 R^2}\int_{-\pi R}^{\pi R} d\tau~\tau^2
\bigg[\sum_{m,n}\frac{(2m^2-5mn+2n^2)/R^2}{(\frac{m^2}
{R^2}+1)}\exp\big(i(n-m)~2 \tau/R\big)\bigg]
\nonumber\\
&&= \frac{1}{16 \pi^2 R^2}\int_{0}^{2 \pi R} d\tau'
~\tau'^2 \sum_{m,n}\frac{(2m^2-5mn+2n^2)/R^2}{(\frac{m^2}
{R^2}+1)}
\nonumber \\
&&\bigg[\exp\big(i(n-m)~\tau'/R\big)+\exp\big(-i(n-m)~\tau'/R\big)\bigg]
\nonumber\\
&&= \int_{0}^{2 \pi R} d\tau' \bigg[ \tau'^2 \delta(-\tau')
~\delta(\tau')-\frac{5}{8}~\tau'^2 \cosh \tau~\delta'(\tau')
\nonumber \\
&&-\frac{1}{2} \coth \pi R ~\tau'^2 \cosh \tau~\delta''(\tau')
\bigg]\,,~~~(\tau'=2 \tau)
\nonumber\\
&&=- \bigg[\frac{5}{8}~\lim_{\epsilon \rightarrow 0}
\frac{\epsilon^2 \cosh (\epsilon)-\epsilon^2 \cosh (-\epsilon)}{2
\epsilon} 
\nonumber \\
&&+ \coth \pi R~\lim_{\epsilon_{1} \rightarrow
0,\epsilon_{2} \rightarrow 0} \big( \sinh \epsilon_{1} \sinh
\epsilon_{2}+\frac{\epsilon_{1}^2+\epsilon_{2}^2}{\epsilon_{1}
\epsilon_{2}} \cosh \epsilon_{1} \cosh \epsilon_{2}\big)\bigg]
\nonumber\\
&&=-2 \coth \pi R\,.
\ea
\ba
&&(3)
\nonumber\\
&&-\alpha~\mbox{Tr} \sum_k\Phi_k\Phi_{-k}=-\frac{\alpha}{4\pi^2
R^2} \int dt_{1} dt_{2}~\mbox{Tr}(\Phi(t_{1})\Phi(t_{2})) \sum_k
\exp\big(i \frac{k}{R}(t_{1}-t_{2})\big) 
\nonumber \\
&&=-\frac{\alpha}{2 \pi R}
\int_{0}^{2 \pi R} dT~\mbox{Tr}\Phi(T)^2\,.
\nonumber\\
\ea Hence, in $O(\alpha)$, the coefficients of the $O(\Phi \Phi)$
terms are given by comparing with the expression \be
\alpha~F_{\alpha 2}(R)\int_{0}^{2 \pi R} dT~\mbox{Tr}
(\frac{1}{2}\Phi(T)^2)+\alpha~\hat F_{\alpha 2}\int_{0}^{2 \pi R}
dT~\mbox{Tr} (\frac{1}{2}\dot \Phi(T)^2)\,, \ee where, after
collecting all the results, \be F_{\alpha 2}(R)=-\frac{1}{\pi
R}+8-4 \coth \pi R\,,~~\hat F_{\alpha 2}=-4 \coth \pi R\,. \ee

\subsection {The terms of order O($\Phi$$\Phi$$\Phi$):}

\ba
(1)~&&\frac{g^2}{6}\mbox{Tr}~\bigg[\sum_{m,n,k}\frac{1}{\big(\frac{m^2}{R^2}+1
\big)\big(\frac{n^2}{R^2}+1\big)\big(\frac{k^2}{R^2}+1\big)}
~\phi_{m-n}\phi_{n-k}\phi_{k-m}\bigg]
\nonumber\\
&&=\frac{g^2}{6} \int \frac{dt_{1} dt_{2} dt_{3}}{(2 \pi R)^3}
~\mbox{Tr}~( \phi(t_{1})\phi(t_{2})\phi(t_{3}))
\nonumber\\
&&\bigg[\sum_{m,n}\frac{\exp\big(i(n-m)~t_{1}/R\big)
\exp\big(i(k-n)~t_{2}/R\big)
\exp\big(i(m-k)~t_{1}/R\big)}{\big(\frac{m^2}{R^2}+1
\big)\big(\frac{n^2}{R^2}+1\big)\big(\frac{k^2}{R^2}+1\big)}
\bigg]\,.
\nonumber\\
\ea

Using redefinition of the variables into the "center of mass" and
the "relative coordinates",
$$
T=\frac{1}{3}(t_{1}+t_{2}+t_{3})\,,
~~~\tau_{1}=(t_{1}-t_{2})\,,~~~~\tau_{2}=(t_{1}-t_{3})\,,
$$
$$
 dt_{1} ~dt_{2} ~dt_{3}
 = J \Big(\frac{t_{1},t_{2},t_{3}}{T,\tau_{1},\tau_{2}}
 \Big)~dT~d\tau_{1} ~d\tau_{2}=dT ~d\tau_{1} ~d\tau_{2}\,.
$$
Considering $\tau_{1}$ and $\tau_{2}$ to be small and keeping the
order $O( \phi^3)$ term, above series could be evaluated as, \ba
&&\frac{g^3}{6}~\mbox{Tr}~\bigg[\sum_{m,n,k}\frac{1}{\big(\frac{m^2}{R^2}+1
\big)\big(\frac{n^2}{R^2}
+1\big)\big(\frac{k^2}{R^2}+1\big)}~\phi_{m-n}\phi_{n-k}\phi_{k-m}\bigg]
\nonumber\\
&&=\frac{g^3}{48 \pi^3 R^3} \int _{0}^{2 \pi R} dT~ \int_{- \pi
R}^{\pi R} d\tau_{1} ~\int_{- \pi R}^{\pi R} d\tau_{2}
\nonumber \\
&&\mbox{Tr}\Big( \phi(T+\frac{\tau_{1}+\tau_{2}}{3})
\phi(T-\frac{2}{3}\tau_{1}+\frac{1}{3}
\tau_{2})\phi(T+\frac{1}{3}\tau_{1}-\frac{2}{3}\tau_{2})\Big)
\nonumber\\
&&\Bigg[\sum_{m,n,k}\frac{\exp \big(i n \tau_{1} /R\big) \exp
\big(-i m \tau_{2}/R\big) \exp \big(i k (\tau_{2}-\tau_{1})/R
\big)} {\big(\frac{m^2}{R^2}+1\big)\big(\frac{n^2}{R^2}+1\big)
\big(\frac{k^2}{R^2}+1\big)} \Bigg]
\nonumber\\
&& \simeq g^3 F_{g3}(R) \int _{0}^{2 \pi R} dT~\mbox{Tr}
(\phi(T)^3/3)  \ea where, \ba && F_{g3}(g,R)
\nonumber\\
&&=\frac{1}{16 \pi^3 R^3}\int_{- \pi R}^{\pi R} d\tau_{1} ~\int_{-
\pi R}^{\pi R} d\tau_{2} \Bigg[\sum_{m,n,k}\frac{ \exp \big(i n
\tau_{1} /R\big) \exp \big(-i m \tau_{2}/R\big) \exp \big(i k
(\tau_{2}-\tau_{1})/R \big)}
{\big(\frac{m^2}{R^2}+1\big)\big(\frac{n^2}{R^2}+1\big)
\big(\frac{k^2}{R^2}+1\big)}  \Bigg]
\nonumber\\
&&=\frac{1}{16 \pi R \sinh^2\pi R}\int_{- \pi R}^{\pi R}
d\tau_{1} ~\int_{- \pi R}^{\pi R} d\tau_{2} ~\cosh(\pi  R -
\tau_1)~\cosh(\pi  R+  \tau_2)
\nonumber \\
&&~\sum_k \frac{\exp \big(i k (\tau_{2}-\tau_{1})/R \big) }
{\big(\frac{k^2}{R^2}+1\big)}
\nonumber\\
&&=\frac{1}{64 \pi R \sinh^2\pi  R} \int_{- 2 \pi R}^{2 \pi
R} d\hat T ~\int_{-2 \pi R}^{2 \pi R} d\hat \tau ~\big(\cosh(2 \pi
 R +\hat \tau)+\cosh  \hat \tau \big)~\sum_k \frac{\exp i k
\hat \tau /R}{\big(\frac{k^2}{R^2}+1\big)}
\nonumber\\
&&=\frac{\pi R}{16  \sinh^3\pi  R } \int_{0}^{2 \pi R} d\hat
\tau \Big[ \big(\cosh(2 \pi  R + \hat \tau)+\cosh  \hat \tau
\big)~\cosh(\pi  R-\hat \tau)
\nonumber\\
&&+\big(\cosh(2 \pi  R - \hat \tau)+\cosh  \hat \tau
\big)~\cosh(\pi  R+ \hat \tau)\Big]
\nonumber\\
&&=\frac{\pi  R}{64  \sinh^3\pi  R (\cosh 2 \pi  R + \cosh 4
\pi  R ) }~~[ 4 \pi  R \big(3 \cosh \pi  R + 2 \cosh3 \pi  R
\nonumber \\
&&+2 \cosh 5 \pi  R + \cosh 7 \pi  R \big) + \sinh \pi  R +
\sinh 3 \pi  R+ \sinh 5 \pi  R+ 2 \sinh 7 \pi  R
\nonumber \\
&&+ \sinh 9 \pi  R ] \ea

\ba
(2)~&&-\alpha g~\mbox{Tr}\Bigg[\sum_{m,n,l}\frac{(4ml-2mn-2ln+n^2)/R^2}
{\big(\frac{m^2}{R^2}+1\big)\big(\frac{l^2}{R^2}+1\big)}
~\phi_{m-l}\phi_{l-n}\phi_{n-m}\Bigg]
\nonumber\\
&&=-\alpha g \int \frac{dt_{1} dt_{2} dt_{3}}{(2 \pi R)^3}
~\mbox{Tr}~(\phi(t_{1})\phi(t_{2})\phi(t_{3}))
\bigg[\sum_{m,n,l}\frac{(4ml-2mn-2ln+n^2)/R^2}{\big(\frac{m^2}{R^2}+1
\big)\big(\frac{l^2}{R^2}+1\big)}
\nonumber\\
&&\exp\big(-i(n-m)~t_{1}/R\big)
\exp\big(-i(l-n)~t_{2}/R\big)
\exp\big(-i(m-l)~t_{1}/R\big)
\bigg]\,.
\nonumber\\
&&=\frac{-\alpha g}{8 \pi^3 R^3} \int _{0}^{2 \pi R} dT~ \int_{- \pi
R}^{\pi R} d\tau_{1} ~\int_{- \pi R}^{\pi R} d\tau_{2}
\nonumber \\
&&\mbox{Tr}\Big( \phi(T+\frac{\tau_{1}+\tau_{2}}{3})
\phi(T-\frac{2}{3}\tau_{1}+\frac{1}{3}
\tau_{2})\phi(T+\frac{1}{3}\tau_{1}-\frac{2}{3}\tau_{2})\Big)
\nonumber\\
&&\Bigg[\sum_{m,n,l}\frac{(4ml-2mn-2ln+n^2)/R^2} {\big(\frac{m^2}{R^2}
+1\big)\big(\frac{l^2}{R^2}+1\big)}
\exp \big(i m \tau_{1} /R\big) \exp
\big(-i n \tau_{2}/R\big) \exp \big(i l (\tau_{2}-\tau_{1})/R
\big) \Bigg]
\nonumber\\
&&=\frac{-\alpha g}{8 \pi^3 R^3} \int _{0}^{2 \pi R} dT~ \mbox{Tr}
\phi(t)^3\int_{- \pi R}^{\pi R} d\tau_{1} ~\int_{- \pi R}^{\pi R}
d\tau_{2} \Bigg[\sum_{m,n,l}\frac{(4ml-2mn-2ln+n^2)/R^2)/R^2}
{\big(\frac{m^2}{R^2} +1\big)\big(\frac{l^2}{R^2}+1\big)}
\nonumber\\
&&\exp \big(i m \tau_{1} /R\big) \exp \big(-i n \tau_{2}/R\big)
\exp \big(i l (\tau_{2}-\tau_{1})/R \big) \Bigg] \ea

Now, \ba &&\frac{-4\alpha g}{8 \pi^3 R^3} \int_{- \pi R}^{\pi R}
d\tau_{1} ~\int_{- \pi R}^{\pi R} d\tau_{2}
\Bigg[\sum_{m,n,l}\frac{ml/R^2} {\big(\frac{m^2}{R^2}
+1\big)\big(\frac{l^2}{R^2}+1\big)}
\nonumber \\
&&\exp \big(i m \tau_{1} /R\big)
\exp \big(-i n \tau_{2}/R\big) \exp \big(i l (\tau_{2}-\tau_{1})/R
\big) \Bigg]
\nonumber\\
&&=\frac{-4\alpha g}{8 \pi^3 R^3} \int_{- \pi R}^{\pi R} d\tau_{1}
~\int_{- \pi R}^{\pi R} d\tau_{2}~2\pi R
\delta(t_2)~\sum_{m,n,l}\frac{ml/R^2} {\big(\frac{m^2}{R^2}
+1\big)\big(\frac{l^2}{R^2}+1\big)} \nonumber\\
&&\Bigg[\exp\big(i m \tau_{1} /R\big) \exp \big(i l
(\tau_{2}-\tau_{1})/R \big)+ \exp\big(-i m \tau_{1} /R\big) \exp
\big(i l (\tau_{2}+\tau_{1})/R \big)\nonumber\\&& \exp\big(i m
\tau_{1} /R\big) \exp \big(-i l (\tau_{2}+\tau_{1})/R \big)
+\exp\big(-i m \tau_{1} /R\big) \exp \big(-i l
(\tau_{2}-\tau_{1})/R \big)\Bigg] \nonumber\\
&&=\frac{2 \alpha g}{\sinh^2 \pi R}\Big[\pi R \cosh 2 \pi R-\sinh
\pi R/2 \Big]\,. \ea

Also,
\ba &&\frac{-\alpha g}{8 \pi^3 R^3} \int_{- \pi R}^{\pi R}
d\tau_{1} ~\int_{- \pi R}^{\pi R} d\tau_{2}
\Bigg[\sum_{m,n,l}\frac{n/R^2} {\big(\frac{m^2}{R^2}
+1\big)\big(\frac{l^2}{R^2}+1\big)}
\nonumber \\
&&\exp \big(i m \tau_{1} /R\big)
\exp \big(-i n \tau_{2}/R\big) \exp \big(i l (\tau_{2}-\tau_{1})/R
\big) \Bigg]
\nonumber\\
&&=\frac{-\alpha g}{8 \pi^3 R^3} \int_{- \pi R}^{\pi R} d\tau_{1}
~\int_{- \pi R}^{\pi R} d\tau_{2}~\Bigg[-2 \pi R
\delta''(\tau_2)\sum_{m,l}\frac{1}{\big(\frac{m^2}{R^2}
+1\big)\big(\frac{l^2}{R^2}+1\big)} \nonumber\\
&&\exp\big(i m \tau_{1} /R\big) \exp \big(i l
(\tau_{2}-\tau_{1})/R \big)+ \exp\big(-i m \tau_{1} /R\big) \exp
\big(i l (\tau_{2}+\tau_{1})/R\big)\nonumber\\&&-2 \pi R
\delta''(-\tau_2)\sum_{m,l}\frac{1} {\big(\frac{m^2}{R^2}
+1\big)\big(\frac{l^2}{R^2}+1\big)} \exp\big(i m \tau_{1} /R\big)
\exp \big(-i l (\tau_{2}+\tau_{1})/R \big)
\nonumber\\
&&+\exp\big(-i m \tau_{1} /R\big) \exp \big(-i l
(\tau_{2}-\tau_{1})/R \big)\Bigg] \nonumber\\
&&=\frac{\alpha g}{4\sinh^2 \pi R}\int_{- \pi R}^{\pi R} d\tau_{1}
\cosh(\pi R+\tau_1)\cosh(\pi R-\tau_1) 
\nonumber \\
&&\lim_{\epsilon \to 0,
\delta \to 0}\big(
(\cosh(\epsilon+\delta)-\cosh(\epsilon-\delta))/\epsilon \delta
\big)
\nonumber\\
&&=\frac{ \alpha g}{4\sinh^2 \pi R}\Big[\pi R \cosh 2 \pi R+\sinh
\pi R/2 \Big] \,.\ea

The last but one term can be evaluated as,
\ba &&\frac{-2\alpha g}{8 \pi^3 R^3} \int_{- \pi R}^{\pi R}
d\tau_{1} ~\int_{- \pi R}^{\pi R} d\tau_{2}
\Bigg[\sum_{m,n,l}\frac{mn/R^2} {\big(\frac{m^2}{R^2}
+1\big)\big(\frac{l^2}{R^2}+1\big)}
\nonumber \\
&&\exp \big(i m \tau_{1} /R\big)
\exp \big(-i n \tau_{2}/R\big) \exp \big(i l (\tau_{2}-\tau_{1})/R
\big) \Bigg]
\nonumber\\
&&=\frac{-2\alpha g}{8 \pi^3 R^3} \int_{- \pi R}^{\pi R} d\tau_{1}
~\int_{- \pi R}^{\pi R} d\tau_{2}~\Bigg[-2\pi R
\delta'(\tau_2)\sum_{m,l}\frac{1} {\big(\frac{m^2}{R^2}
+1\big)\big(\frac{l^2}{R^2}+1\big)} \nonumber\\
&&\exp\big(i m \tau_{1} /R\big) \exp \big(i l
(\tau_{2}-\tau_{1})/R \big)+ \exp\big(-i m \tau_{1} /R\big) \exp
\big(i l (\tau_{2}+\tau_{1})/R \big)\nonumber\\&&-2\pi R
\delta'(-\tau_2)\sum_{m,l}\frac{1} {\big(\frac{m^2}{R^2}
+1\big)\big(\frac{l^2}{R^2}+1\big)} \exp\big(i m \tau_{1} /R\big)
\exp \big(-i l (\tau_{2}+\tau_{1})/R \big)
\nonumber\\
&&+\exp\big(-i m \tau_{1} /R\big) \exp \big(-i l
(\tau_{2}-\tau_{1})/R \big)\Bigg] \nonumber\\
&&=\frac{\alpha g}{\sinh^2 \pi R} \int_{- \pi R}^{\pi R} d\tau_{1}
\sinh(\pi R+\tau_1)\sinh(\pi R-\tau_1) \lim_{\epsilon \to 0}\sinh
\epsilon/\epsilon
\nonumber\\
&&=\frac{ \alpha g}{2\sinh^2 \pi R}\Big[\pi R \cosh 2 \pi R-\sinh
\pi R/2 \Big] \,.\ea

Thus combining all the terms,

\be -\alpha
g~\mbox{Tr}\Bigg[\sum_{m,n,l}\frac{(4ml-2mn-2ln+n^2)/R^2}
{\big(\frac{m^2}{R^2}+1\big)\big(\frac{l^2}{R^2}+1\big)}
~\phi_{m-l}\phi_{l-n}\phi_{n-m}\Bigg] =F_{\alpha g 3} \alpha g
\int_0^{2 \pi R} dt \mbox{Tr} \phi^2(t)/3\,, \ee where, \be
F_{\alpha g 3}=(\frac{13}{4} \frac{\pi R}{\sinh^2 \pi R}\cosh \pi
R-\frac{11}{4} \frac{\sinh 2 \pi R}{2})\,. \ee

\subsection {The terms of order O($A$$A$):}
(1)
\be
-\mbox{Tr}\Bigg[\sum_{n,k}\frac{1}{\frac{n^2}{R^2}+1}~A_kA_{-k}\Bigg]
=\frac{1}{2} \coth \pi R \int_{0}^{2 \pi R} dt ~\mbox{Tr} (A^2(t))
\ee

\ba
(2)&&-\mbox{Tr}\Bigg[ \frac{1}{2!}\sum_{m,n}\frac{(m^2+2mn+n^2)/R^2}
{\big(\frac{m^2}{R^2}+1
\big)\big(\frac{n^2}{R^2}+1\big)}~A_{m-n}A_{n-m}\Bigg]
\nonumber \\
&&=-\frac{1}{8 \pi^2 R^2}\int_{0}^{2\pi R}~dt_1~dt_2
\mbox{Tr}(A(t_1)A(t_2))~\sum_{m,n}\frac{(m^2+2mn+n^2)/R^2}
{\big(\frac{m^2}{R^2}+1
\big)\big(\frac{n^2}{R^2}+1\big)}~\exp(im(t_2-t_1)/R)
\nonumber\\
&&~\exp(in(t_1-t_2)/R)
\nonumber\\
&&\simeq-\frac{1}{4 \pi^2 R^2}\int_{0}^{2 \pi
R}~dT~\mbox{Tr}(A^2(T))~\int_{0}^{\pi R}
d\tau~\sum_{m,n}\frac{(m^2+2mn+n^2)/R^2} {\big(\frac{m^2}{R^2}+1
\big)\big(\frac{n^2}{R^2}+1\big)}
\nonumber \\
&&\big(\exp(i2m\tau/R)~\exp(-i2n\tau/R)+
\nonumber\\
&&\exp(-i2m\tau/R)~\exp(i2n\tau/R)+
\exp(i2m\tau/R)~\exp(i2n\tau/R)+\exp(-i2m\tau/R)~\exp(-i2n\tau/R)\big)
\nonumber\\
&&=-4 \coth \pi R~\int_{0}^{2 \pi R}~dT ~\mbox{Tr}
(A^2(T))-\frac{1}{2 \sinh^2 \pi R}~\int_{0}^{\pi  R}d\tau \big(2
\sinh(\pi R-\tau)\sinh(\pi R +\tau)+
\nonumber\\
&&\sinh(\pi R-\tau)\sinh(\pi R-\tau)+\sinh(\pi R+\tau)\sinh(\pi R
+\tau)\big)~\int_{0}^{2 \pi R}~dT ~\mbox{Tr}(A^2(T))
\nonumber\\
&&=-\Big(4 \coth \pi R+\frac{1}{2}~\pi R~(1+\coth^2\pi
R)-\frac{1}{2}\coth \pi R-\frac{\pi R}{2 \sinh^2 \pi
R}\Big)~\int_{0}^{2 \pi R}~dT ~\mbox{Tr}(A^2(T)) \ea Hence, the
terms of O($A$$A$) are given by, \be G_2(R) \int dt~ \mbox{Tr}
A^2(t)\,, \ee where, \be G_2(R)=(-3 \coth \pi R-\frac{1}{2}~\pi
R~(1+\coth^2\pi R)+\frac{\pi R}{2 \sinh^2 \pi R})\,. \ee

\subsection {The terms of order O($A$$\Phi$$\Phi$):}

\ba (1) &&i\frac{g^2}{2}~\mbox{Tr}\Bigg[\sum_{m,n,l}\frac{(m+l)/R}
{\big(\frac{m^2}{R^2}+1\big)\big(\frac{n^2}{R^2}+1\big)
\big(\frac{l^2}{R^2}+1\big)}~A_{m-l}\phi_{l-n}\phi_{n-m}\Bigg] 
\nonumber \\
&&=i\frac{g^2}{2} \int \frac{dt_{1} dt_{2} dt_{3} }{(2 \pi
R)^3}~\mbox{Tr}\big( A(t_{1}) \Phi(t_{2})\Phi(t_{3})\big)
\nonumber\\
&&\Bigg[\sum_{m,n,l}\frac{(m+l)/R}
{\big(\frac{m^2}{R^2}+1\big)\big(\frac{n^2}{R^2}+1\big)
\big(\frac{l^2}{R^2}+1\big)}
\nonumber \\
&&\exp \big(i l (t_{1}-t_{2})/R \big)
\exp \big(i n (t_{2}-t_{3})/R\big) \exp \big(i m (t_{3}-t_{1})/R
\big) \Bigg]
\nonumber\\
\ea Now we will use the following redefinition of the variables
into the "center of mass" and "relative coordinates",
$$
T
=\frac{1}{3}(t_{1}+t_{2}+t_{3})\,,
~~~\tau_{1}=(t_{1}-t_{2})\,,~~~~\tau_{2}=(t_{1}-t_{3})\,,
$$
$$
 dt_{1} ~dt_{2}
 ~dt_{3}= J \Big(\frac{t_{1},t_{2},t_{3}}{T,\tau_{1},\tau_{2}}
 \Big)~dT~d\tau_{1} ~d\tau_{2}=dT ~d\tau_{1} ~d\tau_{2}\,.
$$
Considering $\tau_{1}$ and $\tau_{2}$ to be small and neglecting
the terms of the order $O(A(t)   \Phi(t)^2)$, $O(A(t) \dot
\Phi(t)^2)$, $O(A(t) \dot \Phi(t) \Phi(t))$, and $O(A(t) \Phi(t)
\dot \Phi(t))$, and keeping terms of the form $O(A(t) [ \Phi(t),
\dot \Phi(t)])$ only, above series could be evaluated as, \ba
&&i\frac{g^2}{2}~\mbox{Tr}\Bigg[\sum_{m,n,l}\frac{(m+l)/R}
{\big(\frac{m^2}{R^2}+1\big)\big(\frac{n^2}{R^2}+1\big)
\big(\frac{l^2}{R^2}+1\big)}~A_{m-l}\phi_{l-n}\phi_{n-m}\Bigg]
\nonumber\\
&&=i \frac{g^2}{8 \pi^3 R^3} \int _{0}^{2 \pi R} dT~ \int_{- \pi
R}^{\pi R}  d\tau_{1} ~\int_{- \pi R}^{\pi R} d\tau_{2}
~\mbox{Tr}\Big( A(T+\frac{\tau_{1}+\tau_{2}}{3})
\Phi(T-\frac{2}{3}\tau_{1}+\frac{1}{3}\tau_{2})
\Phi(T+\frac{1}{3}\tau_{1}-\frac{2}{3}\tau_{2})\Big)
\nonumber\\
&&\Bigg[\sum_{m,n,l}\frac{(m+l)/R}
{\big(\frac{m^2}{R^2}+1\big)\big(\frac{n^2}{R^2}+1\big)
\big(\frac{l^2}{R^2}+1\big)} \exp \big(i l \tau_{1} /R\big)
\exp \big(-i m \tau_{2}/R\big)
\exp \big(i n (\tau_{2}-\tau_{1})/R \big) \Bigg]
\nonumber\\
&& \simeq i \frac{g^2}{8 \pi^3 R^3} \int _{0}^{2 \pi R} dT~\mbox{Tr}
A(T) \big[\Phi(T),\dot \Phi(T \big] \int_{- \pi R}^{\pi R} d\tau_{1}
~\int_{- \pi R}^{\pi R} d\tau_{2}  \big(\frac{\tau_{1}-\tau_{2}}{3}\big)
\nonumber\\
&&\Bigg[\sum_{m,n,l}\frac{(m+l)/R}
{\big(\frac{m^2}{R^2}+1\big)\big(\frac{n^2}{R^2}+1\big)
\big(\frac{l^2}{R^2}+1\big)} \exp \big(i l \tau_{1} /R\big)
\exp \big(-i m \tau_{2}/R\big)
\exp \big(i n (\tau_{2}-\tau_{1})/R \big) \Bigg]
\nonumber\\
&&=\frac{g^2}{12} \big(\pi R \cosh 2\pi R-\frac{1}{2}\sinh 2\pi R \big)\,,
\ea
Where,
\ba
&&\int_{- \pi R}^{\pi R} d\tau_{1} ~\int_{- \pi R}^{\pi R} d\tau_{2}
\big(\frac{\tau_{1}-\tau_{2}}{3}\big)
\Bigg[\sum_{m,n,l}\frac{(m+l)/R}
{\big(\frac{m^2}{R^2}+1\big)\big(\frac{n^2}{R^2}+1\big)
\big(\frac{l^2}{R^2}+1\big)}
\nonumber \\
&&\exp \big(i \frac{l \tau_{1}}{R}\big) \exp \big(-i \frac{m \tau_{2}}{R}\big)
\exp \big(i \frac{n (\tau_{2}-\tau_{1})}{R} \big) \Bigg]
\nonumber\\
&&=\frac{1}{3} \int_{0}^{\pi R} d\tau_{1} \int_{0}^{\pi R}
d\tau_{2}\sum_{m,n,l}\frac{(m+l)/R}
{\big(\frac{m^2}{R^2}+1\big)\big(\frac{n^2}{R^2}+1\big)
\big(\frac{l^2}{R^2}+1\big)}
\Bigg[ (\tau_{1}-\tau_{2})\Big\{\exp \big(i \frac{l \tau_{1}}{R}\big)
\exp \big(-i \frac{m \tau_{2}}{R}\big)
\nonumber\\
&&\exp \big(i \frac{n (\tau_{2}-\tau_{1})}{R} \big)
-\exp \big(-i \frac{l \tau_{1}}{R}\big) \exp \big(i \frac{m
\tau_{2}}{R}\big) \exp \big(-i \frac{n (\tau_{2}-\tau_{1})}{R}
\big)\Big\}  + (\tau_{1}+\tau_{2})\Big\{\exp
\big(i \frac{l \tau_{1}}{R}\big)
\nonumber\\
&&\exp \big(i \frac{m \tau_{2}}{R}\big) \exp
\big(-i \frac{n (\tau_{2}+\tau_{1})}{R} \big)
-\exp \big(-i \frac{l \tau_{1}}{R}\big) \exp
\big(-i \frac{m \tau_{2}}{R}\big) \exp \big(i
\frac{n (\tau_{2}+\tau_{1})}{R} \big)\Big\} \Bigg]
\nonumber\\
&&=-i\frac{2 }{3}\frac{\pi^3 R^3 \cosh \pi R}
{\sinh^3\pi R} \int_{0}^{\pi R} d\tau_{1}
\int_{0}^{\pi R} d\tau_{2}
\Big(\tau_{1} \sinh 2\tau_{1} \cosh 2\tau_{2}-\tau_{2}
\sinh 2\tau_{2} \cosh 2\tau_{1}\Big)
\nonumber\\
&&=-i\frac{2}{3} \pi^3 R^3 \coth^2 \pi R \big(\pi R \cosh 2\pi
R-\frac{1}{2}\sinh 2\pi R \big)\,.
\ea
\ba
(2) &&-i\alpha~\mbox{Tr}\Bigg[\sum_{m,n,l}
\frac{(2m^2l+4ml^2+m^2n-2l^2n-2m^3-mnl)/R^3}
{\big(\frac{m^2}{R^2}+1\big) \big(\frac{l^2}{R^2}+1\big)}
~A_{m-l}\phi_{l-n}\phi_{n-m}\Bigg]
\nonumber\\
&&=-i \alpha \int  \frac{dt_{1} dt_{2} dt_{3} }{(2 \pi R)^3}
~\mbox{Tr}\big( A(t_{1}) \Phi(t_{2})\Phi(t_{3})\big) \Bigg[\sum_{m,n,l}
\frac{(2m^2l+4ml^2+m^2n-2l^2n-2m^3-mnl)/R^3}
{\big(\frac{m^2}{R^2}+1\big)\big(\frac{l^2}{R^2}+1\big)}
\nonumber\\
&&\exp \big(i l (t_{1}-t_{2})/R \big) \exp \big(i n (t_{2}-t_{3})/R\big)
\exp \big(i m (t_{3}-t_{1})/R \big)\Bigg]
\nonumber\\
&&= -\frac{i \alpha}{8 \pi^3 R^3} \int _{0}^{2 \pi R} dT~
\int_{- \pi R}^{\pi R} d\tau_{1} ~\int_{- \pi R}^{\pi R}
d\tau_{2} ~\mbox{Tr}\Big( A(T+\frac{\tau_{1}+\tau_{2}}{3})
\Phi(T-\frac{2}{3}\tau_{1}+\frac{1}{3}\tau_{2})\Phi(T+\frac{1}{3}
\tau_{1}-\frac{2}{3}\tau_{2})\Big)
\nonumber\\
&&\Bigg[\sum_{m,n,l}\frac{(2m^2l+4ml^2+m^2n-2l^2n-2m^3-mnl)/R^3}
{\big(\frac{m^2}{R^2}+1\big) \big(\frac{l^2}{R^2}+1\big)}
\exp \big(i \frac{l \tau_{1}}{R}\big) \exp
\big(-i\frac{ m \tau_{2}}{R}\big)
\exp \big(i \frac{n (\tau_{2}-\tau_{1})}{R} \big) \Bigg]
\nonumber\\
&&\simeq -\frac{i \alpha}{8 \pi^3 R^3} \int _{0}^{2 \pi R}
dT~\mbox{Tr}A(T) \big[\Phi(T),\dot \Phi(T \big]~ \int_{- \pi
R}^{\pi R} d\tau_{1} ~\int_{- \pi R}^{\pi R} d\tau_{2} \big(
\frac{\tau_{1}-\tau_{2}}{3}\big)
\nonumber\\
&&\Bigg[\sum_{m,n,l}\frac{(2m^2l+4ml^2+m^2n-2l^2n-2m^3-mnl)/R^3}
{\big(\frac{m^2}{R^2}+1\big) \big(\frac{l^2}{R^2}+1\big)}
\exp \big(i \frac{l \tau_{1}}{R}\big) \exp
\big(-i\frac{ m \tau_{2}}{R}\big)
\exp \big(i \frac{n (\tau_{2}-\tau_{1})}{R} \big) \Bigg]
\nonumber\\
\ea

Now contribution of the different terms on the above sum can be
evaluated as, \ba (2.a)~&&-\frac{i \alpha}{24 \pi^3 R^3} ~\int_{-
\pi R}^{\pi R} d\tau_{1}~\int_{- \pi R}^{\pi R} d\tau_{2}
\big(\tau_{1}-\tau_{2}\big)
\Bigg[\sum_{m,n,l}\frac{(2m^2l+4ml^2)/R^3}
{\big(\frac{m^2}{R^2}+1\big) \big(\frac{l^2}{R^2}+1\big)}
\exp \big(i \frac{l \tau_{1}}{R}\big) \nonumber\\
&&\exp \big(-i\frac{ m \tau_{2}}{R}\big)
\exp \big(i \frac{n }{R}(\tau_{2}-\tau_{1}) \big) \Bigg]
\nonumber\\
&&=-i\frac{4\alpha}{3} \int_{0}^{\pi R} d\tau_{1}~ \tau_{1}
~\delta(\tau_{1})\sum_{l} \frac{l /R~\sinh l\tau_{1}/R}{((l^2/R^2)+1)}
\nonumber\\
&&=0\,.
\ea

\ba
(2.b)~&&-\frac{i \alpha}{24 \pi^3 R^3} ~\int_{- \pi R}^{\pi R}
d\tau_{1}~\int_{- \pi R}^{\pi R} d\tau_{2} \big(\tau_{1}-\tau_{2}\big)
\Bigg[\sum_{m,n,l}\frac{(2m^2n-2l^2n)/R^3}
{\big(\frac{m^2}{R^2}+1\big) \big(\frac{l^2}{R^2}+1\big)}
\exp \big(i \frac{l \tau_{1}}{R}\big) \nonumber\\
&&\exp \big(-i\frac{ m \tau_{2}}{R}\big)
\exp \big(i \frac{n }{R}(\tau_{2}-\tau_{1}) \big) \Bigg]
\nonumber\\
&&=\frac{-i \alpha}{24 \pi^3 R^3} \int_{0}^{\pi R} d\tau_{1}
\int_{0}^{\pi R} d\tau_{2}\sum_{m,n,l}\frac{m^2n/R^3}
{\big(\frac{m^2}{R^2}+1\big) \big(\frac {l^2}{R^2}+1\big)}
\Bigg[ (\tau_{1}-\tau_{2})\Big\{\exp \big(i n (\tau_{2}-\tau_{1})/R\big)
\nonumber\\
&&\Big(\exp \big(i l \tau_{1}/R\big)\exp \big(-i m \tau_{2} / R \big)
-2 \exp \big(i m \tau_{1}/R\big)\exp \big(-i l \tau_{2}/ R \big)\Big)
\nonumber\\
&&-\exp \big(-i n (\tau_{2}-\tau_{1})/R \big) \Big(\exp \big(-i l
\tau_{1} /R \big)\exp \big(i m \tau_{2}/R \big)
-2 \exp \big(-i m \tau_{1}/R\big)\exp \big(i l \tau_{2}/R\big)\Big)
\Big\}
\nonumber\\
&&+ (\tau_{1}+\tau_{2})\Big\{\exp \big(-i n (\tau_{2}+\tau_{1})/R\big)
\Big(\exp \big(i l \tau_{1}/R\big)\exp \big(i m \tau_{2}/R\big)
-2 \exp \big(i m \tau_{1}/R\big)\exp \big(i l \tau_{2}/R\big)\Big)
\nonumber\\
&&-\exp \big(i n (\tau_{2}+\tau_{1})/R\big) \Big(\exp \big(-i l
\tau_{1}/R\big)\exp \big(-i m \tau_{2}/R\big)
-2 \exp \big(-i m \tau_{1}/R\big)\exp \big(-i l \tau_{2}/R\big)\Big)
\Big\} \Bigg]
\nonumber\\
&&=\frac{\alpha}{3 \pi R} \int_{0}^{\pi R} d\tau_{1}~ \tau_{1}
~\Big(\delta'(-\tau_{1})\sum_{l} \frac{\exp (il\tau_1/R)}
{(l^2/R^2)+1}-\delta'(\tau_{1})\sum_{l} \frac{\exp (-il\tau_1/R)}
{(l^2/R^2)+1}\Big)
\nonumber\\
&&=-\frac{2}{3} \alpha \coth \pi R\,.
\ea

\ba
(2.c)~&&-\frac{i \alpha}{12 \pi^3 R^3} ~\int_{- \pi R}^{\pi R}
d\tau_{1}~\int_{- \pi R}^{\pi R} d\tau_{2} \big(\tau_{1}-\tau_{2}\big)
\Bigg[\sum_{m,n,l}\frac{m^3/R^3}
{\big(\frac{m^2}{R^2}+1\big) \big(\frac{l^2}{R^2}+1\big)}
\exp \big(i \frac{l \tau_{1}}{R}\big)
\exp \big(-i\frac{ m \tau_{2}}{R}\big)
\nonumber\\
&&\exp \big(i \frac{n }{R}(\tau_{2}-\tau_{1}) \big) \Bigg]
\nonumber\\
&&=\frac{-i \alpha}{12 \pi^3 R^3} \int_{0}^{\pi R} d\tau_{1}
\int_{0}^{\pi R} d\tau_{2}\sum_{m,n,l}\frac{m^3/R^3}
{\big(\frac{m^2}{R^2}+1\big) \big(\frac {l^2}{R^2}+1\big)}
\Bigg[ (\tau_{1}-\tau_{2})\Big\{\exp \big(i n (\tau_{2}-\tau_{1})/R\big)
\exp \big(i l \tau_{1}/R\big)
\nonumber\\
&&\exp \big(-i m \tau_{2} / R \big)
-\exp \big(-i n (\tau_{2}-\tau_{1})\exp \big(i m \tau_{2}/R\big)
\exp \big(-i l \tau_{1}/ R \big)\Big\}
\nonumber\\
&&+ (\tau_{1}+\tau_{2})\Big\{\exp \big(i l \tau_{1}/R\big)\exp
\big(i m \tau_{2}/R\big)
\exp \big(-i n (\tau_{2}+\tau_{1})/R\big)
\nonumber\\
&&- \exp \big(-i m \tau_{2}/R\big)
\exp \big(-i l \tau_{1}/R\big)\exp \big(i n (\tau_{2}+\tau_{1})/R\big)
\Big\} \Bigg]
\nonumber\\
&&=\frac{-\alpha}{3 \pi R}\sum_l \frac{1}{\frac{l^2}{R^2}+1}
\int_0^{\pi R}d\tau_1 \int_0^{\pi R} d\tau_2 \Big[ (\tau_{1}-\tau_{2})\Big(
\delta(\tau_2-\tau_1) \delta'(-\tau_2) \exp \big(i l \tau_{1}/R\big)
-\delta(-\tau_2+\tau_1) \delta'(\tau_2)
\nonumber\\
&& \exp \big(-i l \tau_{1}/R\big)\Big)
+(\tau_{1}+\tau_{2})\Big(
\delta(-\tau_1-\tau_2) \delta'(\tau_2) \exp \big(i l \tau_{1}/R\big)
\nonumber\\
&&-\delta(\tau_1+\tau_2) \delta'(-\tau_2) \exp \big(-i l
\tau_{1}/R\big)\Big)\Big]=0
\ea

\ba
(2. d)~&&-\frac{i \alpha}{24 \pi^3 R^3}
~\int_{- \pi R}^{\pi R} d\tau_{1}~\int_{- \pi R}^{\pi R}
d\tau_{2} \big(\tau_{1}-\tau_{2}\big)
\Bigg[\sum_{m,n,l}\frac{mnl/R^3}
{\big(\frac{m^2}{R^2}+1\big) \big(\frac{l^2}{R^2}+1\big)}
\exp \big(i \frac{l \tau_{1}}{R}\big)
\exp \big(-i\frac{ m \tau_{2}}{R}\big)
\nonumber\\
&&\exp \big(i \frac{n }{R}(\tau_{2}-\tau_{1}) \big) \Bigg]
\nonumber\\
&&=\frac{-i \alpha}{24 \pi^3 R^3} \int_{0}^{\pi R} d\tau_{1}
\int_{0}^{\pi R} d\tau_{2}\sum_{m,n,l}\frac{mnl/R^3}
{\big(\frac{m^2}{R^2}+1\big) \big(\frac {l^2}{R^2}+1\big)}
\Bigg[ (\tau_{1}-\tau_{2})\Big\{ \exp \big(i l \tau_{1}/R\big)
\exp \big(-i m \tau_{2} / R \big)
\nonumber\\
&&\exp \big(i n (\tau_{2}-\tau_{1})/R\big)
-\exp \big(i m \tau_{2}/R\big)\exp \big(-i l \tau_{1}/ R \big)
\exp \big(-i n (\tau_{2}-\tau_{1})/R\big)\Big\}
\nonumber\\
&&+ (\tau_{1}+\tau_{2})\Big\{\exp \big(i l \tau_{1}/R\big)\exp
\big(i m \tau_{2}/R\big)
\exp \big(-i n (\tau_{2}+\tau_{1})/R\big)
\nonumber\\
&&- \exp \big(-i m \tau_{2}/R\big)
\exp \big(-i l \tau_{1}/R\big)\exp \big(i n (\tau_{2}+\tau_{1})/R\big)
\Big\} \Bigg]
\nonumber\\
&&=\frac{\alpha}{24 \sinh^2 \pi R} \int_0^{\pi R} d\tau_1
\int _0^{\pi R} d\tau_2
~\Big[(\tau_1-\tau_2)\Big(\delta'(\tau_2-\tau_1) (\cosh(2\pi
R+\tau_2-\tau_1)-\cosh(\tau_2+\tau_1))
\nonumber\\
&&-\delta'(\tau_1-\tau_2)(\cosh(2\pi R+\tau_1-\tau_2)
-\cosh(\tau_1+\tau_2))\Big)
\nonumber\\
&&+(\tau_1+\tau_2)\Big(\delta'(-\tau_2-\tau_1) (\cosh(2\pi
R-\tau_2-\tau_1)-\cosh(\tau_2-\tau_1))-
\nonumber\\
&&\delta'(\tau_1+\tau_2)(\cosh(2\pi R+\tau_1+\tau_2)
-\cosh(\tau_2-\tau_1))\Big) \Big]
\nonumber\\
&&=-\frac{\alpha}{\sinh^2 \pi R} \Big(\frac{\pi R}{6}
\cosh \pi R+\frac{1}{12} \sinh \pi R \Big)
\ea

(3)
\ba
&&i\alpha~\mbox{Tr}\sum_{m,n,k}
\bigg\{\frac{(4n-2m)/R}{\frac{m^2}{R^2}+1}
-\frac{(2n-m)/R}{\frac{n^2}{R^2}+1}\bigg\}~A_{m-n-k}\phi_k\phi_{n-m}
\nonumber \\
&&=\frac{i \alpha}{8 \pi^3 R^3} \int_{0}^{2 \pi R}
dt_1~dt_2~dt_3~\mbox{Tr} (A(t_1)\phi(t_2)\phi(t_3))
\sum_{m,n,k}\bigg\{\frac{(4n-2m)/R}{\frac{m^2}{R^2}+1}
-\frac{(2n-m)/R}{\frac{n^2}{R^2}+1}\bigg\}
\nonumber\\
&&\exp(ik(t_1-t_2)/R)~\exp(im(t_3-t_1)/R)~\exp(in(t_1-t_3)/R)
\nonumber\\
&&=\frac{i \alpha}{24 \pi^3 R^3} \int_{0}^{2 \pi R}
dT~\mbox{Tr} (A(T)[\phi(T),\phi(\dot T)])
~\int_{-\pi R}^{\pi R}~d\tau_1~d\tau_2~(\tau_1-\tau_2)
\sum_{m,n,k}\bigg\{\frac{(4n-2m)/R}{\frac{m^2}{R^2}+1}
-\frac{(2n-m)/R}{\frac{n^2}{R^2}+1}\bigg\}
\nonumber\\
&&\exp(ik(\tau_2-\tau_1)/R)~\exp(im\tau_1/R)~\exp(-in\tau_1/R)
\nonumber\\
&&=0\,. \ea As the overall behavior of the function is
proportional to $\int_{-\pi R}^{\pi
R}~d\tau_1~d\tau_2~(\tau_1-\tau_2) \delta(\tau_1-\tau_2)$~,
therefore the contribution vanishes. Similarly the contribution of
the other term with similar sum \be i \alpha~\sum_{m,n,k}
\bigg\{\frac{(4n-2m)/R}{\frac{n^2}{R^2}+1}
-\frac{(2n-m)/R}{\frac{m^2}{R^2}+1}\bigg\}~A_{m-n-k}\phi_{n-m}\phi_k
 \ee
 vanishes also.
Thus, the O$(A[\phi,\dot \phi])$ term is given by

$$
(G_{g3}~g^2+G_{\alpha 3}~\alpha) \int dt~A(t)[\phi(t),\dot \phi(t)]\,,
$$
where,
\ba
&&G_{g3}(R)=\frac{1}{12}(\pi R \cosh 2 \pi R-\frac{1}{2} \sinh 2 \pi R)
\nonumber\\
&&G_{\alpha 3}(R)=-\frac{2}{3} \coth \pi R-\frac{1}
{\sinh^2 \pi R}(\frac {\pi R}{6}\cosh \pi R+\frac{1}{12}\sinh \pi R)\,.
\ea

\subsection {The terms of order O($A^2\Phi^2$)}

\ba
(1).~ &&2\alpha~\mbox{Tr}~\Bigg[\sum_{m,n,l,k}
\frac{(n^2-2nm+4ml-2nl)/R^2}{\big(\frac{m^2}{R^2}+1\big)
\big(\frac{l^2}{R^2}+1\big)}~2\phi_{l-n}\phi_{n-m}A_{m-l-k}A_k\Bigg]
\nonumber\\
&&=\frac{2 \alpha}{(2 \pi R)^4} \int_0^{2 \pi R} dt_1~dt_2
~dt_3~dt_4~\mbox{Tr}\big(A(t_1)\phi(t_2)\phi(t_3)A(t_4) \big)
\sum_{m,n,l,k}\frac{(n^2-2nm+4ml-2nl)/R^2}{\big(\frac{m^2}{R^2}+1\big)
\big(\frac{l^2}{R^2}+1\big)}
\nonumber\\
&&\exp(i\frac{k}{R}(t_3-t_4))~\exp(i\frac{l}{R}(t_3-t_1))
~\exp(i\frac{n}{R}(t_1-t_2))~\exp(i\frac{m}{R}(t_2-t_3))~
\nonumber\\
\ea

Again using the usual redefinition of the variables
into the "center of mass" and "relative coordinates",
$$
T=\frac{1}{4}(t_{1}+t_{2}+t_{3}+t_{4})\,,~~~\tau_{1}
=(t_{4}-t_{1})\,,~~~~\tau_{2}=(t_{4}-t_{2})
~~~~\tau_{3}=(t_{4}-t_{3})\,,
$$
$$
 dt_{1} ~dt_{2} ~dt_{3}~dt_{4}= J \Big(\frac{t_{1},t_{2},
 t_{3},t_{4}}{T,\tau_{1},\tau_{2},\tau_{3}} \Big)~dT
 ~d\tau_{1} ~d\tau_{2}~d\tau_{3}=dT ~d\tau_{1} ~d\tau_{2}~d\tau_{3}\,.
$$
Considering $\tau_{1}$, $\tau_{2}$ and $\tau_{4}$
to be small and keeping terms of the form
$O(A(t)^2\Phi(t)^2)$ only, above series could be evaluated as,
\ba
&&2\alpha~\mbox{Tr}~\Bigg[\sum_{m,n,l,k}
\frac{(n^2-2nm+4ml-2nl)/R^2}{\big(\frac{m^2}{R^2}+1\big)
\big(\frac{l^2}{R^2}+1\big)}~2\phi_{l-n}\phi_{n-m}A_{m-l-k}A_k\Bigg]
\nonumber\\
&&\simeq \frac{2 \alpha}{(2 \pi R)^3}
\int_{-\pi R}^{\pi R}~d\tau_1~d\tau_2
~\sum_{m,n,l}\frac{(n^2-2nm+4ml-2nl)/R^2}{\big(\frac{m^2}{R^2}+1\big)
\big(\frac{l^2}{R^2}+1\big)}~\exp(i\frac{l}{R}(\tau_1))~
\nonumber\\
&&\exp(i\frac{n}{R}(\tau_2-\tau_1))~\exp(-i\frac{m}{R}(\tau_2))~
\int_0^{2 \pi R} dT ~\mbox{Tr} \Big(\phi^2(T) A^2(T)\Big)
\nonumber\\
&&=\frac{4 \alpha}{(2 \pi R)^3}\int_0^{2 \pi R} dT
~\mbox{Tr} \Big(\phi^2(T) A^2(T)\Big)
 \sum_{m,n,l}\frac{(n^2-2nm+4ml-2nl)/R^2}
 {\big(\frac{m^2}{R^2}+1\big)\big(\frac{l^2}{R^2}+1\big)}
\nonumber\\
&& \int_{-\pi R}^{\pi R}~d\hat \tau_1~d\hat \tau_2
~\exp(i\frac{l}{R}(\hat \tau_1+\hat \tau_2))~
\exp(-i\frac{n}{R} 2 \hat \tau_2)~\exp(i\frac{m}{R}(\hat \tau_2-\hat \tau_1))~
\nonumber\\
\ea The terms containing powers of $n$ in the numerator of the sum
inserts higher and higher derivatives of delta function over time
variables and are eventually computed to be zero. The only
non-vanishing contribution comes from the term without $n$ in the
numerator. \ba &&\frac{4~\alpha}{(2 \pi R)^3}\int_0^{2 \pi R} dT
~\mbox{Tr} \Big(\phi^2(T) A^2(T)\Big)
\sum_{m,n,l}\frac{4ml/R^2}{\big(\frac{m^2}{R^2}+1\big)\big(\frac{l^2}{R^2}+1\big)}
\nonumber\\
&&\int_{0}^{\pi R}~d\hat \tau_1~d\hat \tau_2 ~\Big(\exp(i\frac{l}{R}
(\hat \tau_1+\hat \tau_2))
~\exp(-i\frac{n}{R}2\hat \tau_2)~\exp(i\frac{m}{R}(\hat \tau_2-\hat \tau_1))
\nonumber\\
&&+\exp(i\frac{l}{R}(\hat \tau_1-\hat \tau_2))~\exp(i\frac{n}{R}(2\hat \tau_2))
~\exp(-i\frac{m}{R}(\hat \tau_2+\hat \tau_1))+\exp(-i\frac{l}{R}(\hat \tau_1-\hat
\tau_2))
\nonumber\\
&&~\exp(i\frac{n}{R}(-2 \hat \tau_2))
~\exp(i\frac{m}{R}(\hat \tau_2+\hat \tau_1))
+~\exp(-i\frac{l}{R}(\hat\tau_1+\hat \tau_2))~\exp(i\frac{n}{R}(2 \hat \tau_2)
~\exp(-i\frac{m}{R}(\hat \tau_2-\hat \tau_1))\Big)
\nonumber\\
&&=-\frac{16~\alpha}{\sinh^2 \pi R}\int_0^{2 \pi R}
dT ~\mbox{Tr} \Big(\phi^2(T) A^2(T)\Big)
\int_0^{\pi R} d \hat \tau_1 \Big(\sinh
(\pi R-\hat \tau_1) \sinh(\pi R + \hat \tau_1) \Big)
\nonumber\\
&&=-\frac{8~\alpha}{\sinh^2 \pi R}
(\pi R \cosh 2 \pi R-\frac{1}{2} \sinh 2 \pi R)
\int_0^{2 \pi R} dT ~\mbox{Tr} \Big(\phi^2(T) A^2(T)\Big)
\ea

(2)~~Similarly,
\ba
&&-2~\alpha~\mbox{Tr}\Bigg[\sum_{m,n,l,k}
\frac{(2m^2-nm+2ml-nl)/R^2}{\big(\frac{m^2}{R^2}+1\big)
\big(\frac{l^2}{R^2}+1\big)}~(\phi_{n-m}A_{m-l}A_{l-n-k}\phi_k)\Bigg]
\nonumber \\
&&=\frac{2 ~\alpha}{(2 \pi R)^4} \int_0^{2 \pi R}
dt_1~dt_2~dt_3~dt_4~\mbox{Tr}\big(\phi(t_1)A(t_2)A(t_3)\phi(t_4)\big)
\sum_{m,n,l,k}\frac{(2m^2-nm+2ml-nl)/R^2}{\big(\frac{m^2}{R^2}+1\big)
\big(\frac{l^2}{R^2}+1\big)}
\nonumber\\
&&\exp(i\frac{k}{R}(t_3-t_4))~\exp(i\frac{l}{R}(t_2-t_3))
~\exp(i\frac{n}{R}(t_3-t_1))~\exp(i\frac{m}{R}(t_1-t_2))~
\nonumber\\
&&\simeq \frac{2 ~\alpha}{(2 \pi R)^3} \int_0^{2 \pi R}
dT~\mbox{Tr}\big(\phi(T)A(T)A(T)\phi(T)\big)
\int_{-\pi R}^{\pi R} d\tau_1 d\tau_2 \sum_{m,n,l,k}
\frac{(2m^2-nm+2ml-nl)/R^2}{\big(\frac{m^2}{R^2}+1\big)
\big(\frac{l^2}{R^2}+1\big)}
\nonumber\\
&&\exp(-i\frac{l}{R}\tau_2)~\exp(i\frac{n}{R}\tau_1)
~\exp(i\frac{m}{R}(\tau_2-\tau_1))~
\nonumber\\
&&\simeq \alpha \big(\frac{5}{2} \coth \pi R-2 \pi R
(1+\coth^2 \pi R) \big)~\int_0^{2 \pi R} dT~\mbox{Tr}
\big(\phi(T)A(T)A(T)\phi(T)\big)
\ea
In evaluating above expression again we see that the
term containing (any power of) $n$ in the numerator
of the sum is not contributing. Again,
\ba
&&2~ \alpha~ \mbox{Tr} \Bigg[\sum_{m,n,l,k}
\frac{(2l^2-nm+2ml-nl)/R^2}{\big(\frac{m^2}{R^2}+1\big)
\big(\frac{l^2}{R^2}+1\big)}~(A_{m-l}\phi_{l-n}\phi_kA_{n-m-k})\Bigg]
\nonumber\\
&&=\frac{2 ~\alpha}{(2 \pi R)^4} \int_0^{2 \pi R}
dt_1~dt_2~dt_3~dt_4~\mbox{Tr}\big(A(t_1)\phi(t_2)\phi(t_3)A(t_4)\big)
\sum_{m,n,l,k}\frac{(2l^2-nm+2ml-nl)/R^2}{\big(\frac{2m^2}{R^2}+1\big)
\big(\frac{l^2}{R^2}+1\big)}
\nonumber\\
&&\exp(i\frac{k}{R}(t_4-t_3))~\exp(i\frac{l}{R}(t_1-t_2))
~\exp(i\frac{n}{R}(t_2-t_4))~\exp(i\frac{m}{R}(t_4-t_1))~
\nonumber\\
&&\simeq \frac{2 ~\alpha}{(2 \pi R)^3}\int_0^{2 \pi R}
dT~\mbox{Tr}\big(A(T)\phi(T)\phi(T)A(T)\big)
\int_{-\pi R}^{\pi R} d\tau_1 d\tau_2 \sum_{m,n,l,k}
\frac{(2l^2-nm+2ml-nl)/R^2}{\big(\frac{m^2}{R^2}+1\big)
\big(\frac{l^2}{R^2}+1\big)}
\nonumber\\
&&\exp(i\frac{m}{R}\tau_1)~\exp(-i\frac{n}{R}\tau_2)
~\exp(i\frac{l}{R}(\tau_2-\tau_1))\,.
\nonumber\\
\ea
Exchanging $l \leftrightarrow m$ and $\tau_1
\leftrightarrow -\tau_2$, this gives equal and
opposite contribution to the previous expression
and hence all the similar pair of sums
\ba
&&-\alpha\sum_{m,n,l,k}\frac{(2m^2-nm+2ml-nl)/R^2}{\big(\frac{m^2}{R^2}+1\big)
\big(\frac{l^2}{R^2}+1\big)}~(2A_{l-n-k}\phi_k\phi_{n-m}A_{m-l}-\phi_kA_{l-n-k}
\phi_{n-m}A_{m-l})
\nonumber \\
&&+\alpha\sum_{m,n,l,k}\frac{(2l^2-nm+2ml-nl)/R^2}{\big(\frac{m^2}{R^2}+1\big)
\big(\frac{l^2}{R^2}+1\big)}~(2\phi_{l-n}\phi_kA_{n-m-k}A_{m-l}
-\phi_{l-n}A_{n-m-k}\phi_kA_{m-l})\,,
\nonumber\\
\ea
give zero contribution.

\ba
(3)~&&\alpha~\mbox{Tr}\Bigg[ \sum_{m,n,k,k'}
\bigg\{\frac{4}{\frac{m^2}{R^2}+1}+\frac{1}{\frac{n^2}{R^2}+1}\bigg\}
~A_{m-n-k}\phi_k\phi_{k'}A_{n-m-k'}\Bigg]
\nonumber\\
&&=\frac{\alpha}{(2 \pi R)^4} \int_0^{2 \pi R}
dt_1~dt_2~dt_3~dt_4~\mbox{Tr}\big(A(t_1)\phi(t_2)\phi(t_3)A(t_4) \big)
\sum_{m,n,k,k'}\Big( \frac{4}{\frac{m^2}{R^2}+1}+\frac{1}{\frac{n^2}{R^2}+1}\Big)
\nonumber\\
&&\exp(i\frac{k}{R}(t_1-t_2))~\exp(i\frac{k'}{R}
(t_4-t_3))~\exp(i\frac{n}{R}(t_1-t_4))~\exp(i\frac{m}{R}(t_4-t_1))~
\nonumber\\
&&\simeq \frac{5 \alpha}{2} \coth \pi R \int_0^{2 \pi R} dT
~\mbox{Tr}\big(A(T)\phi(T)\phi(T)A(T) \big)
\ea

(4)~~Similraly,
\ba
&&-\alpha~\mbox{Tr}\Big[\sum_{m,n,k,k'}\frac{2}{\frac{m^2}{R^2}+1}
\big(\phi_k A_{m-n-k}\phi_{k'} A_{n-m-k'}+\phi_kA_{n-m-k'}
\phi_{k'}A_{m-n-k}\big)\Big]
\nonumber \\
&&=\frac{-\alpha}{(2 \pi R)^4} \int_0^{2 \pi R}
dt_1~dt_2~dt_3~dt_4~\mbox{Tr}\big(A(t_1)\phi(t_2)A(t_3) \phi(t_4)
\big) \sum_{m,n,k,k'}\Big(\frac{2}{\frac{n^2}{R^2}+1}\Big)
\nonumber\\
&&\Big\{\exp(i\frac{k}{R}(t_2-t_1))~\exp(i\frac{k'}{R}(t_4-t_3))
~\exp(i\frac{n}{R}(t_2-t_4))~\exp(i\frac{m}{R}(t_4-t_2))~
\nonumber\\
&&+\exp(i\frac{k}{R}(t_4-t_1))~\exp(i\frac{k'}{R}(t_2-t_3))
~\exp(i\frac{n}{R}(t_4-t_2))~\exp(i\frac{m}{R}(t_2-t_4))\Big\}
\nonumber\\
&&\simeq -2 \alpha \coth \pi R \int_0^{2 \pi R} dT
~\mbox{Tr}\big(A(T)\phi(T)A(T) \phi(T)\big)
\ea

(5)
\ba
&&-\frac{g^2}{2}~\mbox{Tr}\Big[\sum_{m,n,l,k}
\frac{1}{\big(\frac{m^2}{R^2}+1\big)\big(\frac{n^2}{R^2}+1\big)
\big(\frac{l^2}{R^2}+1\big)}~\phi_{l-n}\phi_{n-m}A_{m-l-k}A_k\Big]
\nonumber\\
&&=-\frac{g^2}{2(2 \pi R)^4}~\int_0^{2 \pi R}
dt_1~dt_2~dt_3~dt_4~\mbox{Tr}\big(A(t_1)\phi(t_2)A(t_3) \phi(t_4) \big)
\sum_{m,n,l,k}
\frac{1}{\big(\frac{m^2}{R^2}+1\big)\big(\frac{n^2}{R^2}+1\big)
\big(\frac{l^2}{R^2}+1\big)}
\nonumber\\
&&\exp(i\frac{k}{R}(t_3-t_4))~\exp(i\frac{l}{R}(t_3-t_1))
~\exp(i\frac{n}{R}(t_1-t_2))~\exp(i\frac{m}{R}(t_2-t_3))~
\nonumber\\
&&\simeq-\frac{g^2}{16 (\pi R)^3}\int_0^{2 \pi R} dT
~\mbox{Tr}\big(A(T)\phi(T)A(T) \phi(T)\big)~\sum_{m,n,l,}
\frac{1}{\big(\frac{m^2}{R^2}+1\big)\big(\frac{n^2}{R^2}+1\big)
\big(\frac{l^2}{R^2}+1\big)}
\nonumber\\
&&\int_{-\pi R}^{\pi R} d\tau_1 d\tau_2~\exp(-i\frac{l}{R}
\tau_2)~\exp(i\frac{n}{R}\tau_1)~\exp(i\frac{m}{R}(\tau_2-\tau_1))~\,,
\nonumber\\
\ea
where,
\ba
&&-\frac{g^2}{16 (\pi R)^3}~\sum_{m,n,l,}
\frac{1}{\big(\frac{m^2}{R^2}+1\big)\big(\frac{n^2}{R^2}+1\big)
\big(\frac{l^2}{R^2}+1\big)}
\int_{-\pi R}^{\pi R} d\tau_1 d\tau_2~\exp(-i\frac{l}{R}
\tau_2)~\exp(i\frac{n}{R}\tau_1)~\exp(i\frac{m}{R}(\tau_2-\tau_1))
\nonumber\\
&&=-\frac{g^2}{8 \pi R \sinh \pi R}~\sum_{m}\frac{1}{\big(\frac{m^2}{R^2}+1\big)}
\int_{0}^{\pi R} d\tau_1 d\tau_2~\big(\sinh (\pi R+\tau_2)
\sinh(\pi R -\tau_1)\exp(i\frac{m}{R}(\tau_2-\tau_1))+
\nonumber\\
&&\sinh (\pi R+\tau_2)\sinh(\pi R +\tau_1)\exp(i\frac{m}{R}(\tau_2+\tau_1))\big)
\nonumber\\
&&=-\frac{g^2}{16 \pi R \sinh \pi R}~\sum_{m}\frac{1}{\big(\frac{m^2}{R^2}+1\big)}
\int_{0}^{\pi R} d\tau_1 d\tau_2~\Big(\big(\cosh (2\pi R+\tau_2-\tau_1)-
\cosh(\tau_1+\tau_2)\big)
\nonumber \\
&&\exp(i\frac{m}{R}(\tau_2-\tau_1))+
\nonumber\\
&&\big(\cosh (2\pi R+\tau_1+\tau_2)-\cosh(\tau_1-\tau_2)\big)
\exp(i\frac{m}{R}(\tau_2+\tau_1))\Big)
\nonumber\\
&&=-\frac{g^2}{8 \pi R \sinh \pi R}~\sum_{m}\frac{1}{\big(\frac{m^2}{R^2}+1\big)}
\int_{0}^{\pi R} d\hat T~\int_{-\pi R/2}^{\pi R/2} d\hat \tau~\Big(\big
(\cosh (2\pi R+2 \hat \tau)-
\cosh(2 T)\big)\exp(i\frac{2 m}{R} \hat \tau)+
\nonumber\\
&&\big(\cosh (2\pi R+2 T)-\cosh(2 \hat \tau)\big)\exp(i\frac{2 m}{R} T)\Big)
\nonumber\\
&&=-\frac{g^2}{8 \pi R \sinh \pi R}~\Big[\int_{-\pi R/2}^{\pi R/2} d\hat
\tau~\big(\pi R~\cosh (2\pi R+2 \hat \tau)-\frac{1}{2}\sinh (2 \pi R)\big)
\sum_{m}\frac{1}{\big(\frac{m^2}{R^2}+1\big)}~\exp(i\frac{2 m}{R} \hat \tau)+
\nonumber\\
&&\int_{0}^{\pi R} d T~\big(\pi R \cos
h (2\pi R+2 T)-\sinh(\pi R) \big)~\frac{\pi R}{\sinh \pi R} \cosh (\pi R- 2T)\Big]
\nonumber\\
&&=-\frac{g^2}{8 \sinh^2 \pi R}~\int_{0}^{\pi R} dT\Big[~\big(\frac{\pi R}{2}
~\cosh (2\pi R+T)-\frac{1}{4}\sinh 2 \pi R\big)\cosh(\pi R-T)+\big
(\frac{\pi R}{2}~\cosh (2\pi R-T)
\nonumber\\
&&-\frac{1}{4}\sinh 2 \pi R\big)\cosh(\pi R+T)+
\big(\pi R \cosh (2\pi R+2 T)-\sinh(\pi R) \big)~ \cosh (\pi R- 2T)\Big]
\nonumber\\
&&=-\frac{g^2}{8 \sinh^2 \pi R} \Big( \pi^2 R^2 \cosh 3\pi R+\frac{\pi R}{8}
\sinh 5 \pi R+
\frac{\pi R}{4} \sinh 3 \pi R 
\nonumber \\
&&- \frac{3}{8} \sinh \pi R + \frac{1}{4}
\sinh^2 2 \pi R-\sinh^2 \pi R\Big)
\nonumber\\
\ea

Collecting all the terms, the O$(A\phi A\phi)$ and the O$(\phi^2 A^2)$
terms are given by
$$
(G_{g4}g^2+G_{\alpha 4}\alpha\int dt~\mbox{Tr}A(t)\phi(t)A(t)\phi(t)+(G'_{g4}g^2
+G'_{\alpha 4}\alpha) \int dt~\mbox{Tr}A^2(t)\phi^2(t)\,,
$$
where,
\ba
&&G_{g4}(R)=-\frac{1}{8 \sinh^2 \pi R} \Big( \pi^2 R^2 \cosh 3\pi R
+\frac{\pi R}{8} \sinh 5 \pi R+
\frac{\pi R}{4} \sinh 3 \pi R - \frac{3}{8} \sinh \pi R
\nonumber\\
&&+ \frac{1}{4} \sinh^2 2 \pi R-\sinh^2 \pi R\Big)
\nonumber\\
&&G'_{g4}(R)=0
\nonumber\\
&&G_{\alpha 4}(R)=-2 \coth \pi R
\nonumber\\
&&G'_{\alpha 4}(R)= \frac{5}{2} \coth \pi R\,.
\ea

\appendix{The Scaling Dimensions}\label{A3}

For $\Lambda^*_1$, \ba &&\Omega_{11}=\frac{4
F_{g3}+3F_{g2}+3F_{g1}+\frac{3}{2}G_{g4}}{\hat
F_{g2}-F_{g1}-F_{g2}-3 G_{g4}-2F_{g3}}\,,
\nonumber\\
&&\Omega_{12}=\frac{3 \hat F_{\alpha 2}+2F_{\alpha g 3}}
{\hat F_{g2}-F_{g1}-F_{g2}-3 G_{g4}-2F_{g3}}\,,
\nonumber\\
&&\Omega_{21}=0\,,
\nonumber\\
&&\Omega_{22}=-2 \Big(\frac{\hat F_{g2}+F_{g2}+F_{g1}-2G_{g3}
+\frac{3}{2}G_{g4}+F_{g3}}
{\hat F_{g2}-F_{g1}-F_{g2}-3 G_{g4}-2F_{g3}}\Big)\,.
\ea

For $\Lambda^*_2$,
\ba
&&\Omega_{11}=\frac{6 F_{\alpha g3}
+6F_{\alpha2}+3G_{\alpha4}-2G_2}{\hat F_{\alpha2}-F_{\alpha2}-3 G_{\alpha4}
+2G_2-2F_{\alpha g3}}\,,
\nonumber\\
&&\Omega_{12}=0\,,
\nonumber\\
&&\Omega_{21}=0\,,
\nonumber\\
&&\Omega_{22}=\frac{6 \hat F_{\alpha2}+2 F_{\alpha2}-8G_{\alpha3}
+3G_{\alpha4}+2 G_2+2F_{\alpha g3}}
{\hat F_{\alpha2}-F_{\alpha2}-3 G_{\alpha4}+2G_2-2F_{\alpha g3}}\,.
\ea


\end{document}